\documentclass[%
 aip,
 amsmath,amssymb,
 preprint,%reprint
]{revtex4-1}

\usepackage{amsmath, amssymb}

\usepackage{physics}
\usepackage[dvipsnames]{xcolor}
\usepackage{amssymb}
\usepackage{ulem}
\usepackage{mathtools} 
\usepackage{comment} 
\usepackage{xspace} 
\usepackage{enumitem}
\usepackage{soul}

\usepackage{etoolbox}

\makeatletter
\def\@email#1#2{%
 \endgroup
 \patchcmd{\titleblock@produce}
  {\frontmatter@RRAPformat}
  {\frontmatter@RRAPformat{\produce@RRAP{*#1\href{mailto:#2}{#2}}}\frontmatter@RRAPformat}
  {}{}
}%
\makeatother

\newcommand{\pref}[1]{(\ref{#1})}
\newcommand{\epref}[1]{Eq.~(\ref{#1})}

\newcommand{\figref}[1]{Fig.~\ref{#1}}

\newcommand{\half}{\frac 12}
\newcommand{\iu}{\mathrm i}

\newcommand{\ie}{\textit{i.e.}}

\newcommand{\eg}{\textit{e.g.}}

\newcommand{\si}[1]{supplementary material#1}

\newcommand{\helc}{\lambda}

\newcommand{\su}{\uparrow}
\newcommand{\sd}{\downarrow}
\newcommand{\Lld}{\mathfrak{L}}
\newcommand{\Rld}{\mathfrak{R}}
\newcommand{\Rl}[2]{R^{(\Lld\Lld)}_{#1 #2, m}}
\newcommand{\Rr}[2]{R^{(\Rld\Rld)}_{#1 #2, m}}

\newcommand{\Tll}[2]{T^{(\Rld\Lld)}_{#1 #2, m}}
\newcommand{\Trl}[2]{T^{(\Lld\Rld)}_{#1 #2, m}}
\newcommand{\Rll}[2]{R^{(\Lld\Lld)}_{#1 #2, m}}
\newcommand{\Rrl}[2]{R^{(\Rld\Rld)}_{#1 #2, m}}
\newcommand{\EF}{E_\mathrm{F}}
\newcommand{\qF}{q_\mathrm{F}}

\newcommand{\spinq}{\frac e{4\pi}}
\newcommand{\s}[1]{\hat\sigma_{#1}}

\newcommand{\revisiondiff}{0}
\newcommand{\revisionnew}{1}

\newcommand{\revisiontype}{\revisionnew}
\newcommand{\revision}[2]{%
\if\revisiontype\revisiondiff
     {\color{Red}\st{#1}}{\color{Green}#2}
\else
     \if\revisiontype\revisionnew
         #2
     \else
         #1
     \fi
\fi}

\begin{document}

\author{Richard Korytár}%
\email{richard.korytar@ur.de}
\affiliation{Department of Condensed Matter Physics, Faculty of Mathematics and Physics, Charles University, Ke Karlovu 5, 121 16, Praha 2, Czech Republic}

\author{Jan M. van Ruitenbeek}
\affiliation{Huygens-Kamerlingh Onnes Laboratory, Leiden University, NL-2333CA Leiden, Netherlands}

\author{Ferdinand Evers}
\affiliation{Institute of Theoretical Physics, University of Regensburg, D-93050 Regensburg, Germany}

\title{Spin conductances and magnetization production in chiral molecular junctions}

\keywords{Chirality-induced spin selectivity, molecular junctions, spintronics}

%\preprint{arXiv:2404.05614 [cond-mat.mes-hall]}

%\begin{tocentry}
%\includegraphics{tocentry_michaeli}\hfill
%\end{tocentry}

\begin{abstract}
Motivated by experimental reports on chirality induced spin selectivity, we investigate a minimal model that allows us to calculate the charge and spin conductances through helical molecules analytically. The spin-orbit interaction is assumed to be non-vanishing on the molecule and negligible in the reservoirs (leads). The band-structure of the molecule features four helical modes with spin-momentum locking that are analogous of edge-currents in the quantum spin Hall effect. 
While charge is conserved and therefore the charge current is independent of where it is measured, reservoirs or molecule, our detailed
calculations reveal that the spin currents in the left and right leads are equal in magnitudes but
with opposite signs (in linear response).
 We predict that transport currents flowing through helical molecules are accompanied by a spin accumulation in the contact region with the same magnetization direction for source and drain. Furthermore, we predict that the spin-conductance can be extracted directly from measuring the (quasi-static) spin accumulation - rather than the spin current itself, which is very challenging to obtain experimentally. 
\end{abstract}

\maketitle

%\pagebreak

\section{Introduction}

Charge currents are routinely measured and analyzed in molecular electronics \cite{Evers2020}. The discovery of a family of phenomena that exhibit chirality induced spin selectivity (CISS) has led to a resurge of interest in spin-related phenomena in this field.\cite{Gohler2011,Naaman2019,Liu2020,RezaSafari2022,RezaSafari2023,Theiler2023} 
Specifically, experiments report a strong correlation between molecular chirality and a preferred spin direction in systems that exhibit a (nominally) very weak spin-orbit interaction. At present, there is no consensus concerning the explanation of many experimental CISS results \cite{Evers2022}. 

Motivated by the CISS phenomena, we here address charge and spin currents in chiral molecular junctions within the framework of a minimal model. As a diagnostic tool of the junction's atomic structure, spin currents offer advantages as compared to charge currents: 
Spin polarized currents can, in principle, be detected in charge-transport measurements within an analyzer-polarizer setup employing, e.g., magnetized leads. However, they are expected to 
manifest themselves only in the non-linear regime and not in the linear charge conductance $G(\mathbf M)$.
This is a consequence of Onsager's reciprocity; its importance for
the theory of the CISS effect was emphasized by Yang, van der Wal and van Wees\cite{Yang2019, Yang2020}.
\footnote{As is well known, due to the Onsager relations magnetotransport experiments based on the polarizer-analyzer setup will not detect spin-currents in the linear transport regime: even if a non-vanishing spin-current is flowing, the charge conductance is independent of the magnetization direction $G(\mathbf{M})=G(-\mathbf{M})$ and therefore insensitive to the flow of spin.}  The spin-conductance, on the other hand, %is itself a proper observable in spatial regions that exhibit spin rotational invariance. It 
is less restricted by Onsager symmetries. It can be inferred, at least in principle, from measuring the pile-up of magnetization, e.g., in source or drain. A brief review of the symmetry properties of transport coefficients is available in the \si{, Sec.~S1}.

These considerations motivated us and other researchers\cite{Wolf2023} to investigate a minimal model for a chiral molecule in the presence of spin-orbit interactions. While Ref.\cite{Wolf2023} has focussed on the effect of contact-enhanced SOI, we adopt the model proposed by Michaeli and Naaman\cite{Michaeli2019}; it exhibits SOI on the entire molecule and has the extra benefit of allowing for an analytical treatment. 
While Michaeli and Naaman have studied the transmission properties, our focus is on (spin) conductances. As one would have expected, the spin conductance turns out to be non-vanishing due to spin-orbit coupling. As a consequence, the transmitted and  reflected currents tend to build up a non-vanishing spin-accumulation near both contacts, source and drain, already in the linear regime\cite{Garcia2023}. In stationary non-equilibrium the magnitude of induced magnetization is likely controlled by spin-relaxation processes. We predict the orientation of the magnetization in 
both contacts to be the same, in agreement with requirements of time-reversal invariance (TRI).\cite{Wolf2023} 

Our work is of potential impact for constructing a molecular machine. 
Since total angular momentum is conserved by the $LS$-coupling, spin-flip processes exert a mechanical torque that can drive an engine. An analogue driving mechanism based on angular transfer has been investigated in Ref. \cite{Korytar2023}.  

%%%%%%%%%%%%%%%%%%%%%%%%%%%%%
\section{Results}
\subsection{Minimal modeling of a helical molecule}
%\subsection{Hamiltonian of electrons bound to a helical tube}

\begin{figure}[t]
\includegraphics[width=.75\columnwidth]{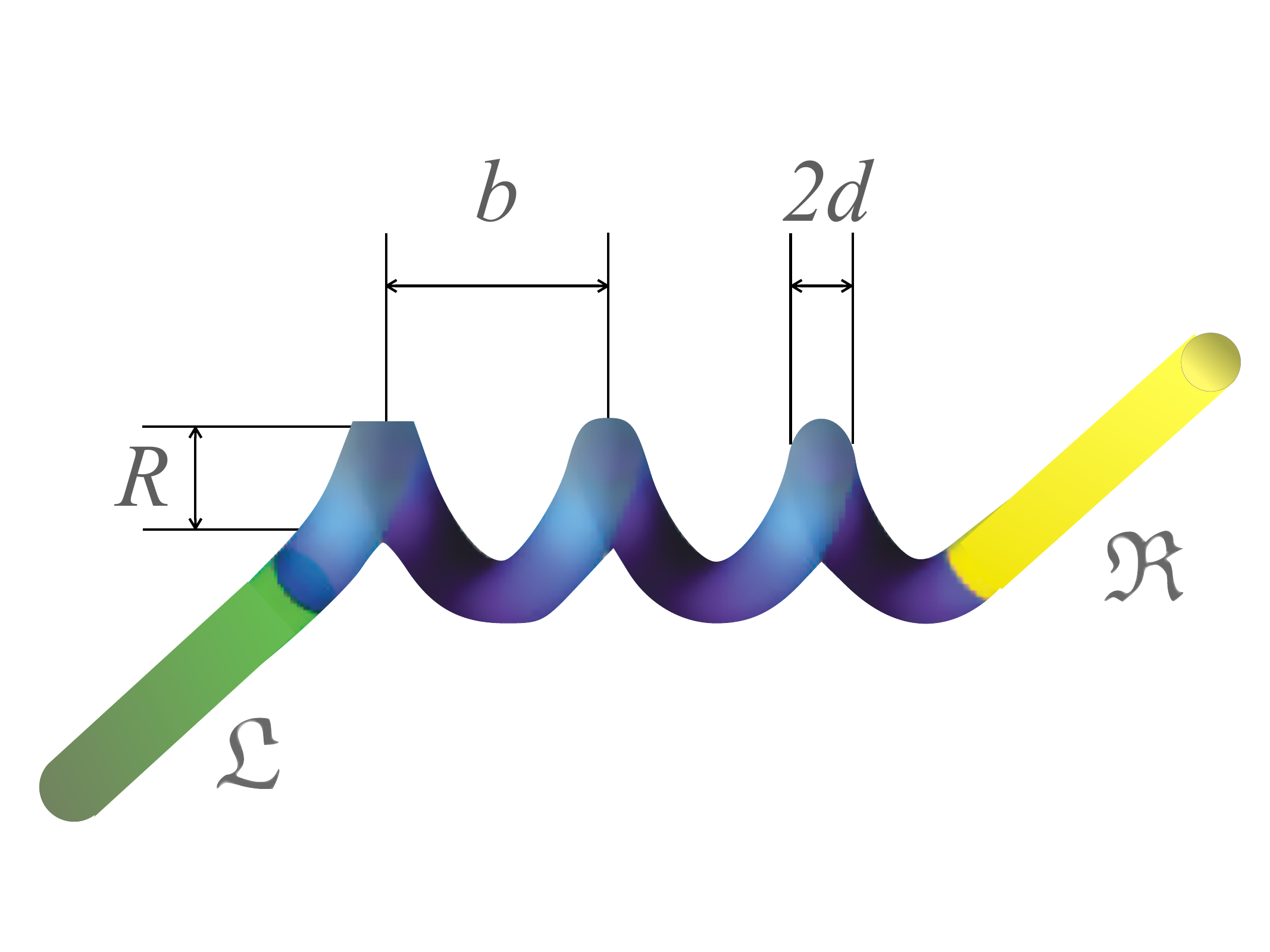}
\includegraphics[width=.6\columnwidth]{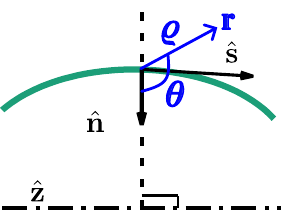}
\caption{\label{fig:hcoord}
Top: Scheme of the molecular junction consisting of a helical
tube (center) and a pair of semi-infinite straight tubes, left and right.
The radius $R$, pitch $b$, thickness $2d$ are indicated.
Bottom: Illustration of helical coordinates $s,\varrho,\theta$:
Green line shows part of a helix that evolves along the $z$
axis (dash-dotted line) with unit vector $\hat{\mathbf z}$.
At any given point on the helix, the unit vector
$\hat{\mathbf n}$ \revision{is perpendicular to $\hat{\mathbf z}$ and 
the unit tangent vector $\hat{\mathbf s}$}{, the unit tangent vector $\hat{\mathbf s}$
and $\hat{\mathbf z}$, are perpendicular (indicated by the small rectangle)}. In helical coordinates,
$s$ is the distance along the helix; $\theta$ and $\varrho$ are polar
coordinates \revision{}{of a vector $\vb{r}$ lying} in a plane normal to $\hat{\mathbf s}$; $\theta$ measures
from $\hat{\mathbf n}$.}
\end{figure}

Following Michaeli and Naaman\cite{Michaeli2019} we consider electrons bound to a long tube.  The left and right side of the tube are of cylindrical shape and represent (semi-infinite) reservoirs; the central region in between takes a helical shape to mimic a chiral molecule, see Fig. 
\ref{fig:hcoord}.

 % We (re-)derive an effective Hamiltonian that defines the minimal model. 
%Further, the helicity of the tube will be denoted by $\helc = \pm 1$.
The Schr\"odinger equation that describes the free motion of a particle inside the helical tube reads
\begin{equation}
\Bigl[-\frac{\hbar^2\grad ^2}{2m_e} + V_\mathrm H(\mathbf r) +
\frac{\iu \hbar^2}{4m_e^2c^2}
\vb*{\sigma}\cdot \grad  V_\mathrm H(\mathbf r) \times
\grad {}
\Bigr]\Psi(\vb r) = E\Psi(\vb r),
\label{eq:sche}
\end{equation}
where ${\mathbf r}$ is the position vector in three dimensions; $V_\mathrm H(\mathbf r)$ denotes
an effective single-particle potential that confines the electron to the tube (see Fig. 
\ref{fig:hcoord}) with tube radius $d$, helical radius $R$ and pitch $b$. The third term 
%on the right-hand 
in the square brackets
represents the spin-orbit interaction (SOI).

In the limit of small $d/\tilde R$ the electronic wavefunction
in the central region
is tightly bound to the helix and the quasi one-dimensional nature of the model becomes manifest; here, $\tilde R = \sqrt{(2\pi R)^2 + b^2}>0$ denotes the distance covered when completing one helical turn. In this limit, the longitudinal and transverse motion approximately decouple and the problem simplifies.  
Following this idea, a systematic expansion of \epref{eq:sche}
in $d/\tilde R$ 
has been performed by Michaeli and Naaman\cite{Michaeli2019} and keeping only leading order terms a minimal model has been derived. 
%%%%%%%%%%%%%%%%%%%%%%%%%%%%%

\subsection{Limit of a narrow tube}
\subsubsection{Wavefunction factorization}
%In the quasi one-dimensional limit 
The wavefunction $\Psi(s,\varrho,\theta)$ can be 
conveniently expressed in a local cylindrical coordinate system:
$s$ denotes the longitudinal coordinate (distance along the helix);
the motion in the plane normal to the
tangential vector $\hat{\mathbf s}$ is described by the radial coordinate $\varrho$, which denotes
normal distance from the center line of the tube (Fig. \ref{fig:hcoord}),
and $\theta$, which denotes the corresponding angular coordinate. 

For simplicity, we will assume local rotational invariance in the sense 
 $V_\mathrm H(\vb r) = V_\mathrm H(\varrho)$.
In the small $d/\tilde R$ limit, the longitudinal and transverse motion nearly
decouple,\cite{Michaeli2019} and the wavefunction 
factorizes,\footnote{The details of the confinement $V_\mathrm H(\varrho)$
are not relevant for the main conclusions of our work. Focusing on the lowest-energy eigenstates, we adopt a harmonic approximation $V_\mathrm H(\varrho) 
\approx \half \frac{\hbar^2}{m_{\mathrm e}d^4}\varrho^2$: the resulting wavefunctions $\Phi_{N,m}(\theta,\varrho)$ are eigenstates of a 2D harmonic oscillator with
$m$-independent eigen-energies $E_N$
and $m$ restricted to $ -N, -N+2, \ldots, N-2, N$.
}
$\Psi(s,\varrho,\theta) \approx \bar\Psi_m(s)\Phi_{N,m}(\varrho)\, e^{\iu m\,\theta}$.
The quantum number $N \in \mathbb N_0$ governs the nodal structure of 
the wavefunction in the radial direction.
Due to the stipulated rotational invariance, the $\hat{\mathbf s}$-component
of the angular momentum, $-\iu\hbar\partial/\partial \theta$,
 is a good quantum number; we call it $\hbar m$ and
$m\in \mathbb Z$.
Finally, owing to the presence of SOI, $\bar\Psi_m(s)$ represents a two-component
spinor.  

In leading order, \epref{eq:sche} turns to
$$\hat H_{N,m}(s)\bar\Psi_m(s) = E{\bar\Psi_m(s)},$$ where
\begin{equation}
\hat H_{N,m}(s) = E_{N}\ - \frac{\hbar^2}{2m_\mathrm e}
\qty( \pdv[2]{}{s}\, +\, \iu \frac{2 m b}{\tilde R^2} \pdv{}{s}) 
\, +\, \hat H_\text{soi},
\label{eq:hs}
\end{equation}
with $E_N$ being the energy of the radial and angular motion. The second term
on the right is the kinetic energy operator of the longitudinal
motion. The third term couples momentum along ${s}$ with the
$\hat{\vb{s}}$-component of the orbital angular momentum, $\hbar m$.
The origin of this term is geometric, it arises due to a non-zero pitch
$b$ of the helix.  The last term, $\hat H_\text{soi}$, is the
SOI, which we simplify further in the following.

\subsubsection{Spin-orbit coupling term} In the narrow-tube limit an expression
for the spin-orbit term was given in the Ref.~\cite{Michaeli2019},
\begin{equation}
\label{eq:soi}
\hat H_\text{soi} = \kappa m \Bigl[ \s{x}\sin\qty(\frac{2\pi\lambda s}{\tilde R})
-\s{y}\cos\qty(\frac{2\pi\lambda s}{\tilde R}) -\s{z} \frac{b}{2\pi R}
\Bigr]
\end{equation}
where $\kappa = \lambda (\hbar^4 R)/(4 m_e^3 c^2 d^4\tilde R)$ and
$\helc = +1$ ($-1$) for a right (left) handed helix.

This expression adopts a transparent form on invoking a Cartesian
representation
\begin{equation*}
\hat{\mathbf s}(s) =
-\frac{2\pi R}{\tilde R} \sin \qty(\frac{\helc 2\pi s}{\tilde R}) \hat{\vb{x}}
+\frac{2\pi R}{\tilde R} \cos \qty(\frac{\helc 2\pi s}{\tilde R}) \hat{\vb{y}} 
+ \frac{b}{\tilde R} \hat{\vb{z}}.
\end{equation*}
With ${\mathbf L}= m\hbar\hat{\mathbf s}$ it is easy to see
that 
\begin{equation}
\label{eq:LS}
\hat H_\text{soi} = \beta\, \mathbf L\cdot \vb*{\sigma},\quad
 \beta = -\frac{\helc\hbar^3 }{8\pi m_e^3 c^2 d^4}.
\end{equation}
As $s$ increases along the helix, $\vb{L}(s)$
precesses around the $z$-direction.
This spatial dependency of ${\mathbf L}(s)$ is the main entry point of helicity into
the quasi-one-dimensional model. The precession of ${\vb{L}}$ invites an
analogy with a magnetic moment $\vb*{\mu}$
 in a rotating magnetic field with ``Zeeman energy''
$|\vb*{\mu}\cdot \vb B|= \hbar\beta m$ and
${\bf B}=B\qty[\sin (\Omega s) \hat{\vb{x}} -\cos(\Omega s)\hat{\vb{y}} + B_z/B\, \hat{\vb{z}}]$;
 $s$ being the effective ``time'' and oscillation frequency $\Omega =2\pi/\helc\tilde R$.
Motivated by this observation, in the following paragraph we adopt a
transformation to the rotating frame \cite{Abragam1961} in the next subsection.

\subsubsection{Non-Abelian gauge transformation}

To highlight the conservation of angular momentum in this model, we 
rewrite \eqref{eq:LS} introducing ladder operators 
\begin{align}
\hat H_\text{soi} &= \beta
\left[ \s{+}\hat L_-(s) + \hat\sigma_{-}\hat L_+(s) 
+ \s{z} \hat L_z
\right] 
\label{e3} \\
&=  \kappa m
\left[ \s{+}\,\iu e^{-\iu \Omega s}
- \s{-}\,\iu e^{\iu\, \Omega s} - \s{z} \frac{b}{ 2\pi R}
\label{e5}
\right]
\end{align}
where 
$\s{\pm}=(\s{x}\pm\iu \s{y})/2$
 and $\hat L_{\pm} = \hat L_x
\pm \iu\hat L_y$. 

In the representation \pref{e5}, flipping the spin boosts the  
momentum {along the tube axis}
by a reciprocal lattice vector $2\pi/\tilde R$. Alternatively to $s$, one can label the helical motion with an angle $\delta=\helc 2\pi s/\tilde R$ that takes unique values on the entire real axis;
moving up one pitch implies a change of $\delta\mapsto\delta {\pm} 2\pi$. From this perspective, the phase-factors in \eqref{e5} boost the angular $\delta$-dependency of the spinor
$\Psi(\delta \frac{\helc\tilde R}{2\pi},\varrho,\theta)$ by 
an extra factor $e^{\iu\delta}$,
and in this sense angular momentum is conserved.

While the phase factor in \eqref{e5} accounts for the conservation of the total angular momentum and therefore is crucial, it also obstructs an easy analytical solution of the model because it is not translationally invariant. 
The Hamiltonian considerably simplifies after  
a gauge transformation, $\bar\Psi_m(s) \mapsto e^{\iu \helc \s{z} \pi s/\tilde R}\bar\Psi_m(s)$,
and accordingly for operators.\footnote{
All operators transform as
\begin{align*}
%\label{eq:gtr}
\hat A \mapsto e^{\iu{\helc} \s{z} s\pi/\tilde R} \hat A e^{-\iu \helc\s{z} s\pi/\tilde R}.
\end{align*}
To understand the action of the exponential operator on the spin raising and lowering operators in Eq. \eqref{e5}, one recalls that $\s{+}$ ($\s{-}$) annihilates spin up (spin down) states. Therefore the transformed operator 
\begin{align*}
e^{\iu{\helc} \s{z} s \pi/\tilde R} \s{+} e^{-\iu \helc\s{z} s\pi/\tilde R}
\end{align*}
annihilates a spin-up state - as it would without
being transformed - or equip a spin-down state with two times the same phase factor, i.e. $e^{2\iu \helc s\pi/\tilde R}$. As a consequence, the transformation indeed removes the phase factor of $\s{+}$ seen in Eq. \eqref{e5}. 
}

The gauge-transformed Hamiltonian of the spin-orbit interaction
follows from \eqref{e5},
\begin{align}
\hat H_\text{soi} \mapsto \hat H_\text{soi} &=  -\kappa m 
\left[ \s{y} + \s{z} \frac{\helc b}{2\pi R}
\right].
\label{e6}
\end{align}
Whilst the term $\propto\s{z}$ in \eqref{e5} remains invariant under the transformation, 
the rotating transverse components of the SOI in \eqref{e5} collapse onto a
single spatial direction, $\s{y}$. The latter direction does not follow from the geometry
of the helix, but from our gauge-choice that associates the identity operator with the position $s=0$.

%%%%%%%%%%%%%%%%%%%%%%%%%%%%%%%%%%%%%%%%%%%%%%%%%%%%%%%%%%%%%%%%%%%%%%%%%%%%%%%%%%%%%%%

%%%%%%%%%%%%%%%%%%%%%%%%%%%%%%%%%%%%%%%%%%%%%%%%%%%%%%%%%%%%%%%%%%

\subsubsection{Minimal model Hamiltonian}
The effect of the gauge transformation on the longitudinal momentum operator
is given by a spin-dependent ``boost'' $-\iu{\partial}/{\partial s} \mapsto -\iu{\partial}/{\partial{s}}  - \frac{\helc  \pi}{\tilde R}
\s{z} $. Along with \pref{e6}
the application of the gauge transformation to the 
\epref{eq:hs}
leads to
\begin{equation}
\label{eq:hams}
\hat H_{N,m}(s) \mapsto \hat H_{N,m}(s)
= \frac{\hbar^2}{2m_e}
\qty(-\iu \pdv{}{s} - \helc \frac \pi{\tilde R} \s{z} + \frac{mb}{\tilde R^2})^2
-\kappa m 
\left( \s{y} + \s{z} \frac{\helc b}{2\pi R}
\right)
\end{equation}
% -\kappa  m \s{y} - \lambda\frac{\kappa b}{2\pi R}m\s{z}
where we discarded $E_N$ and a constant $|m|$-dependent energy shift.
The analysis simplifies upon introducing dimensionless variables
$\hat H_{N,m} =  \frac{\hbar^2 \pi^2}{m_e\tilde R^2}\hat H'_{N,m}$
and $s = s'\frac{\tilde R}\pi$, leading to
\begin{equation}
\hat H'_{N,m}(s') = \half\left(
 -{\iu}\frac{\partial}{\partial s'}
-\helc\s{z}
 + \tilde\gamma m \right)^2
 - \tilde\kappa m \left(\helc\s{y} + \s{z} \frac{b}{2\pi R}\right).
 \label{e7}
\end{equation}
In the above expression,
$\tilde \kappa = \frac{\hbar ^2 R\tilde R }{4{\pi^2} m_e^2 d^4c^2} > 0$
and $\tilde\gamma {=}\frac {b}{\pi \tilde R}$.

The $\tilde \gamma$-term corresponds to a momentum shift and thus represents a
"synthetic" vector potential. It can formally be removed from $\hat H'_{N,m}$
by dressing the wavefunctions with a gauge-factor $e^{\iu \tilde\gamma m
s'}$, \cite{Geyer2020} and therefore leaves the spectrum invariant.  If
$\tilde\gamma m$ has a natural  interpretation as a vector potential, the
term proportional to $b/R$ in the second line of \eqref{e7} is the
corresponding "synthetic Zeeman term". Notice that vector potential,
$\tilde\gamma m$, and the Zeeman term change sign under $m\mapsto -m$;
they do not break TRI because the full Hamiltonian sums
over all $m$.

\subsection{Dispersion relation}
The model Hamiltonian \eqref{e7} is straightforward to analyze. 
We  focus on the effect of spin flips and therefore discard synthetic fields.
Translational invariance of the Hamiltonian \eqref{e7} suggests a representation in Fourier space 
\begin{equation} \label{e8}
\hat h_m(k)= \half \left( k - \helc\s{z}\right)^2 -
 \helc\tilde\kappa m\s{y}. 
\end{equation}
The model \eqref{e8} exhibits a two-band dispersion $E_{m,\alpha}(k)$, where $\alpha=\pm 1$; see 
\si{, Sec.~S2}, for explicit expressions. 
\begin{figure}[t]
\includegraphics[width=\columnwidth]{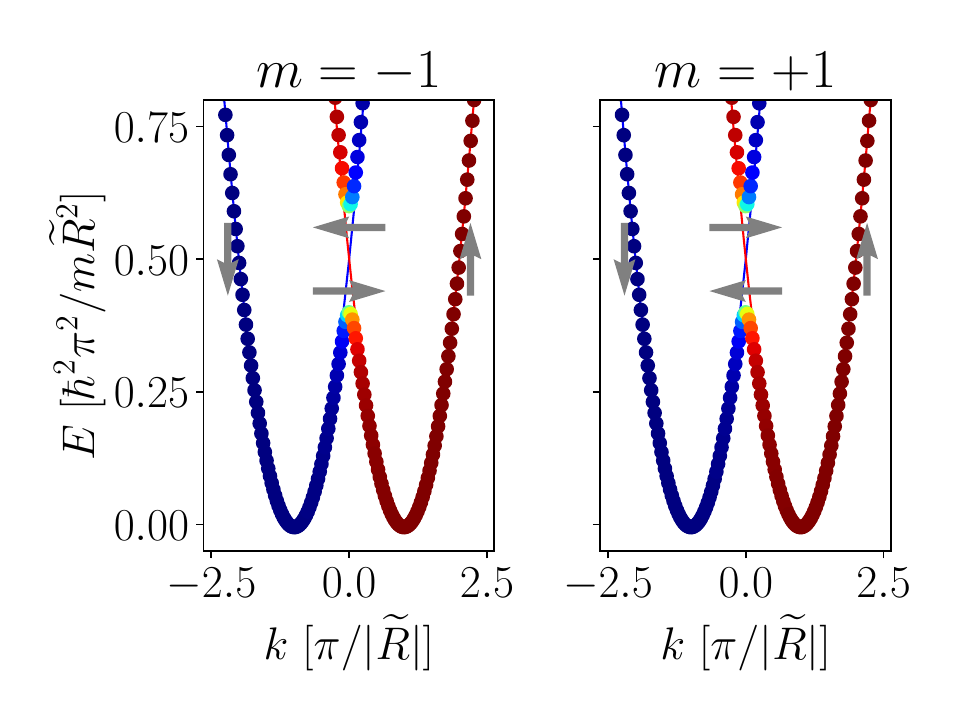}
\caption{\label{fig:dispersion}
Dispersion relation of the helical tube for $m=+1$ and $-1$ is shown with dots.
Color indicates the expectation value of $\s{z}$ (red=+1, blue=-1,
green=0).  Grey arrows represent the expectation value of spin in the
$yz$-plane [$\s{y}$ ($\s{z}$) is the horizontal 
(vertical) component, respectively].
 For comparison, the thin solid lines depict the dispersion of a
straight tube.  The horizontal spin-dependent shift of the parabol{\ae} reflects the non-Abelian gauge transformation. Parameters: $\tilde\kappa = 0.1, \helc = -1$. The
eigen-energies are $m$-independent.}
\end{figure}

\figref{fig:dispersion} shows the dispersion law; the horizontal shift of the two parabol{\ae},
red and blue, reflects the non-Abelian gauge transformation.
The spin-orbit term is effective only at the crossing of parabol{\ae}, where it opens a spin-orbit gap,
$2|m|\tilde\kappa$, as is easily confirmed by degenerate perturbation theory at the crossing point $k=0$.
Chirality-induced spin-selective phenomena are expected to be strong in this region of energies.

In the gap region there remain four %($2\cdot2$) 
ungapped bands, - a factor of two for spin and angular momenta each, - 
where the spin projection and the sign of the velocity, 
$\dd E_{m,\alpha}(k)/\dd k$, are locked. In all  four bands, the projection of
spin onto the direction of 
velocity equals $-\helc\hbar/2$ ("spin-momentum locking").
Such states are termed \textit{helical}, in full analogy to the helicity concept for edge states in the quantum spin Hall effect\cite{Koenig2008}.
The band-structure arising from \pref{e8}
was also discussed in the context of Rashba quantum wires
\cite{Sanchez2008,Streda2003} and chiral carbon 
nanotubes \cite{Marganska2018}, although the physical origin of the terms of the 
Hamiltonian was different from our situation. Specifically, the gap of the 
minmal model \eqref{e6} opens due to the SOI term,  
$-\helc\tilde\kappa m\s{y}$, while in Rashba wires and nanotubes
such term originated from a transverse magnetic field.
%

%\section{Results}
\subsection{Spin and charge transport}

\subsubsection{Molecule bound to straight tubes}
To facilitate transport studies, we attach two straight tubes, Fig. \ref{fig:hcoord}, which serve as reservoirs.
Formally, the reservoirs are included by extending the model \eqref{e7}, so that $\tilde\gamma,\tilde\kappa=0$ if $s'<0$ or $s'>L$. 
Importantly, the non-Abelian gauge transformation restores translational invariance also after attaching leads provided that it is performed in the reservoirs, too. Similarly, the synthetic vector potential, $\tilde\gamma(s') m$, can still be removed by applying a (non-local) gauge factor $e^{\iu m \tilde\Gamma(s')}$, with $\tilde \gamma(s') = \partial_{s'} \Tilde{\Gamma}(s')$.

\subsubsection{Basic definitions}
A finite bias drop $\mu_\Lld -\mu_\Rld = eV$ causes the flow of charge
and spin, described by the charge current $I(V)$ and spin currents
$I^{(\Lld)}_i(V)$, $I^{(\Rld)}_i(V)$ in each lead and each spatial
direction $i=x,y,z$.
 Charge conductance $G$ and spin conductances $G^{(\Lld)}_i,G^{(\Rld)}_i$
 are defined by the linear response relations
\begin{subequations}
\begin{align}
I(V) &= GV + \mathcal O\qty( V^2)\\
I^{(\Lld)}_i(V) &= G^{(\Lld)}_iV + \mathcal O\qty( V^2) \\
I^{(\Rld)}_i(V) &= G^{(\Rld)}_iV + \mathcal O\qty( V^2)
\end{align}
\end{subequations}

Due to local charge conservation, the charge current is well defined and,
in particular, independent of where it is measured along the current flow. 
In contrast, spin is not locally conserved in the presence of spin-orbit coupling.
It is only in the leads of the extended model where (longitudinal) spin-currents are well defined observables. 
Notice that due to the loss of spin-conservation in the central region, $0<s<L$,
spin-currents in left- and right- reservoirs, $I^{(\Lld)}_i$ and $I^{(\Rld)}_i(V)$, 
may differ in a quasi-stationary non-equilibrium situation.

We adopt here the following sign convention for the spin currents: The
$I^{(\Lld)}_i$ measures spin entering the junction from the left ($\Lld$) contact
and $I^{(\Rld)}_i(V)$ measures spin exiting the junction into the
right ($\Rld$) contact. This is fully analogous to the definition of the charge
current.

\subsubsection{Transmissions and Landauer Formul{\ae}}
The Landauer 
formalism  relates conductances to the spectral transmission probabilities: for the transmission from $\Rld$
to $\Lld$, $\Trl{\sigma}{\sigma'}(E)$, and vice versa, $\Tll{\sigma}{\sigma'}(E)$, and to the corresponding reflection amplitudes, $\Rll{\sigma}{\sigma'}(E)$ and $\Rrl{\sigma}{\sigma'}(E)$; see Ref.~\cite{Jacquod2012} for an overview. Here, the right (left) subscripts and superscripts of the transmission  probability label the quantum numbers of an incoming (outgoing) wave, respectively. 
For example, $\Tll{\su}{\sd}$ denotes the
probability to transmit an electron with spin down from the left lead
to the right one while flipping its spin in eigenchannel $m$.
The charge conductances and the $z$-components of the spin conductances 
are given by
\begin{subequations}
\begin{align}
G &= \frac{e^2}h\sum_m \
\sum_{\sigma\sigma'}\Tll{\sigma}{\sigma'}(\EF), \label{e12}\\
G_z^{(\Rld)} &= \spinq \sum_m
\sum_{\sigma\sigma'}\sigma\Tll{\sigma}{\sigma'}(\EF),
\label{eq:gzr}\\
G_z^{(\Lld)} &= \spinq \sum_m
\sum_{\sigma\sigma'}(-\sigma)\Rll{\sigma}{\sigma'}(\EF)
\label{eq:gzl0}
\\
 &= \spinq \sum_m
\sum_{\sigma\sigma'}\sigma\Trl{\sigma}{\sigma'}(\EF),
\label{eq:gzl}
\end{align}
\label{eq:coeffs}
\end{subequations}
where $\EF$ indicates the Fermi energy (the \si{, Sec.~S5}, offers
a standard derivation within the scattering formalism).
The expression for $G$ chosen here emphasizes transmission of
all spin species from left to right.
From that expression the right spin conductance is obtained
by multiplication by $\frac 1e\frac\hbar2\sigma $ in the right lead.
The left spin conductance expressed in the \epref{eq:gzl0} can
then be understood as due to the reflected flux in the $\Lld$-lead.
Particle conservation (unitarity, see \si{, Sec.~S4}),
leads to the equivalent
form, \epref{eq:gzl}.

\subsubsection{Transport Results}

We address the transport problem by calculating the scattering matrix using conventional wavefunction matching. In this process, angular momentum $m$ matches at the two interfaces: it is conserved in the scattering process; for further details see \si.

As %an exemplary case,
example we focus on $N=1$, so that $m=\pm 1$. We further continue to ignore the effect of the synthetic fields. In passing, we briefly mention that their effect is to assign a preferred spin direction, up or down, to a given angular momentum $m$. Hence, they will result in a  circulating (transverse) spin current. In the following our focus is on the longitudinal currents.  
%{\color{red} It seems there is no difference between $m=1$ and $m=-1$ in the transmission coefficients. But isn't this an artefact of dropping $\tilde\gamma$ in \eqref{e8}?  Or does this result from TRI? }
%

\begin{figure}[t]
\includegraphics[width=1.1\columnwidth]{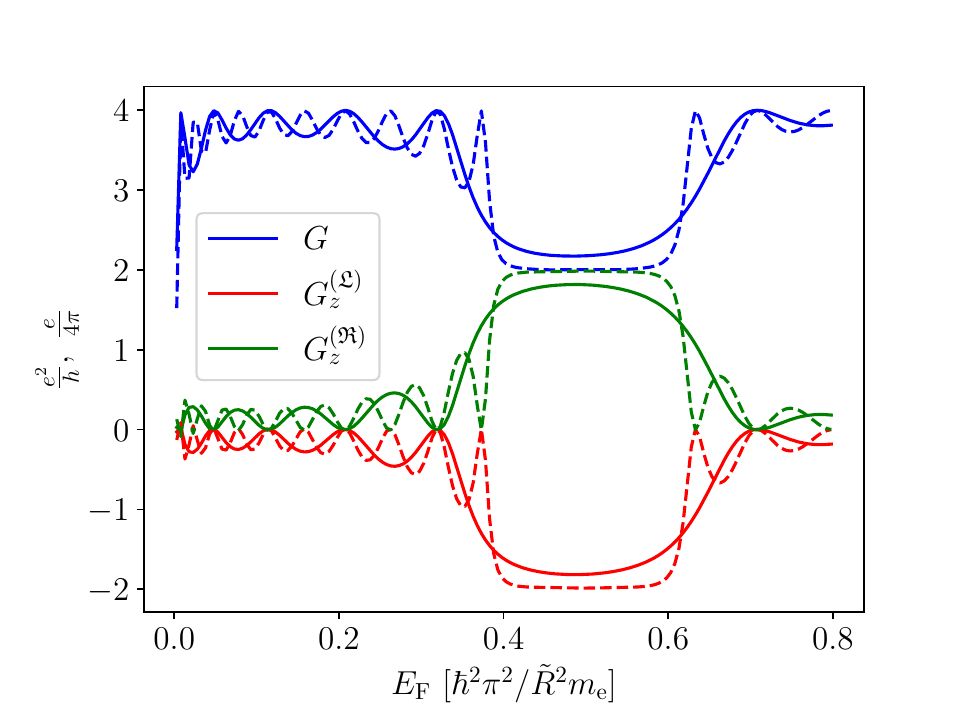}
\caption{\label{fig:coefficients}
Charge conductance and the $z$-components of the spin
 conductances in the Michaeli model as a function of 
the Fermi energy $\EF$. 
\revision{Solid (dashed) lines are for a helix with 3 (dashed) and 6 (solid) turns}{Solid
(dashed) lines are for a helix with 3 (6) turns, respectively}. Parameters are the same as in \figref{fig:dispersion}, with a spin-orbit
gap between 0.4 and 0.6.}
\end{figure}

The energy dependence of the resulting conductances for charge and spin in this model is displayed 
in \figref{fig:coefficients}. We offer a few comments: 
\begin{itemize}[leftmargin=0.5cm,topsep=0em]
\setlength\itemsep{-0.0em}
\item With $\EF$ well outside the spin-orbit gap, the effect of spin-orbit coupling is small and translational invariance is hardly broken. In this case, back-scattering is weak and the (charge) conductance reaches a maximum
of four conductance quanta reflecting two spin and two orbital ($m=\pm 1$) channels. 
\item 
In the off-gap regime mesoscopic oscillations are visible. The oscillation frequency (in $\EF$) is seen to decrease with the inverse length, $L^{-1}$ and therefore we assign the oscillations to Fabry-P\'erot interference. 
\item 
With $\EF$ inside the 
spin-orbit gap, backscattering inside the the wire is suppressed and the Fabry-P\'erot oscillations quickly die out.
\item In the in-gap regime, electrons can tunnel via two evanescent modes
that result from the two `gapped' bands.
Accordingly, the charge conductance in the middle of the gap approaches 2 conductance quanta from above. 
\item 
The absolute value of the spin conductances in either one of the leads %reaches 
approaches $2\,\spinq$.
This can be understood from \figref{fig:dispersion}: in the spin-orbit gap
 the right (or left)
moving modes  have identical spin regardless of $m$, \ie{} there are two
channels %differentiated 
distinguished only
by the orbital angular momentum.
\item 
Importantly, the spin conductances are non-zero for energies even 
far from the spin-orbit gap, where the oscillations peak at $\approx
0.2\, \spinq$. It can be shown that that $G_x^{(\Lld,\Rld)}(\EF)=0$ for
any $\EF$ since the Bloch functions have zero average $\s{x}$.
Moreover,  $G_y^{(\Lld,\Rld)}=0$ because the expectation values of 
$\s{y}$ change sign along with the sign change of $m$ (see \figref{fig:dispersion}),
while the lead Hamiltonian is $m$-independent. In other words,
the $y$ components of Bloch waves exactly cancel upon the
summation over $m$.
\item 
Remarkably, the $\Lld$ and $\Rld$ spin conductances
differ only up to a sign.
By combining TRI and left-right reflection it
is possible to derive relations between scattering matrix elements
that lead to the exact identity $G_z^{(\Rld)} = - G_z^{(\Lld)}$
 at any energy. (See \si{, Sec.~S4,S5} for the details of this symmetry analysis.)
In realistic molecular junctions a symmetry of couplings to the left
and right leads can not be expected. We show in the 
Sec.~\ref{sec:barrier} that in this more general situation, the magnitude of 
$G_z^{(\Rld)}$ and $G_z^{(\Lld)}$ no longer is the same, while the sign is still opposite
due to TRI.
%\vspace{0em} 
\end{itemize}
\noindent 

%%%%%%%%%%%%%%%%%%%%%%%%%%%%
\subsection{Qualitative understanding of spin transport}
%%%%%%%%%%%%%%%%%%%%%%%%%%%%%%%%%%%%%
\subsubsection{A frequent misconception \label{sec:misconception}} 

We begin the discussion by addressing a common misconception of the minimal model: 
\figref{fig:dispersion} is frequently interpreted as predicting a nonzero spin current even in equilibrium. 
Indeed, 
\figref{fig:dispersion} seems to suggest that for a Fermi level situated in the spin-orbit gap there is an excess of spin-up right movers
(velocity $\dd E_{m,\alpha}(k)/\dd k\, >\, 0$) over spin-down right movers.
Taken at face value this observation would imply that due to the wire connecting $\Lld$ and $\Rld$ reservoirs, 
%either reservoir 
both reservoirs 
would become magnetized
at equilibrium. 
Clearly, such a transport of magnetization is violating the second law of thermodynamics.
The vanishing of equilibrium spin currents  
is a rigorous consequence of 
unitarity, see \si{, Sec.~S5}. The fact that spin transport can not be
derived from the band-structure alone is known in the field of two-dimensional
materials, see Sec.~4.1 of Ref.~\cite[Sec.~4.1]{Bercioux2015} for a review 
and a discussion in the context of CISS in Ref.~\cite{Yang2019}.
The paradox will be resolved on a more intuitive level in the subsequent discussion.

\subsubsection{Long helix limit} 
We consider $\helc=-1$ and the limit %$L\rightarrow\infty$, 
$L\gg d,R$
and Fermi energy, $\EF$, inside the spin-orbit gap. In this limit, at observation points
deep inside the one-dimensional wire all up-spin fermions flow one direction, while all down-spin fermions flow into the other direction, see \figref{fig:dispersion}. We infer that $\Trl{\sd}{\sd},\Tll{\su}{\su}\simeq 1$; there is no spin-flip inside the wire. On the other hand, deep inside either reservoir the spin-orbit gap vanishes. Therefore, spin-up and spin-down currents flow alike in either direction. It is easy to see that both limits match if  $\Rll{\su}{\sd}, \Rrl{\sd}{\su}\simeq 1$. 
\begin{figure}[t]
\centering
\includegraphics[width=1.0\columnwidth]{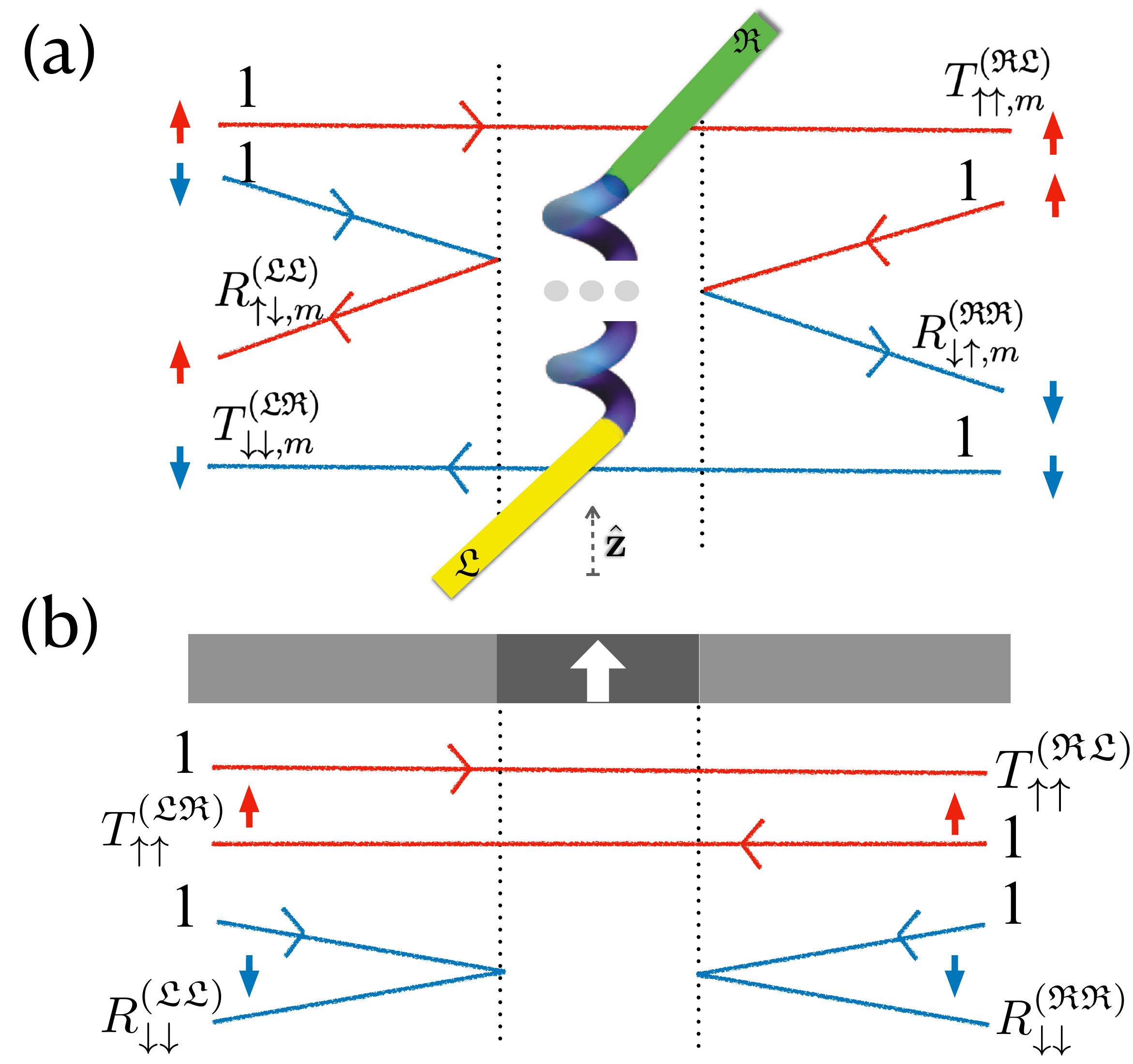}
\caption{\label{fig:scattering}
Schematic representation of scattering processes in the helical
junction (a) and a ferromagnetic polarizer (b). Panel (a) applies in the limit of a long
helix (helicity $\helc=-1$) for energies in the spin-orbit gap. Red (blue) color indicates spin up (down) in the sense
of the helical axis. The red line on the top indicates that the electron with spin $\su$ coming from
the left $(\Lld)$ lead \revision{(normalized to probability 1)}{} is transmitted
and its spin is conserved. \revision{}{"1" indicates that the incident wave is normalized to unit density; the
spin conserving transmission probability $\Tll{\su}{\su}$ is close to one.}
If the incoming electron has spin
$\sd$, it is reflected with a spin-flip due to TRI.
Panel (b) illustrates scattering off an ideal ferromagnetic polarizer that only transmits spin up electrons. }
\end{figure}

The situation is illustrated in \figref{fig:scattering}(a). The central region of the junction 
%in \figref{fig:scattering}(a) 
shows counter-propagating electrons with opposite spin, as suggested by \figref{fig:dispersion}. 
These modes are properly interpreted as carrying a conductance quantum for charge, because charge is conserved inside the wire; hence, in Fig. \ref{fig:coefficients} the conductance is seen to be $G\approx e^2/h$ per angular-momentum channel inside the gap. These same modes are not properly interpreted in terms of spin-conductances, however, because spin is not conserved inside the wire. 

 The propagation pattern of current channels in the leads is also displayed in \figref{fig:scattering}(a); the matching condition at the interface shown there follows from bulk limits. Since spin is conserved inside the reservoirs, the channel pattern may be interpreted in terms of spin currents.
 \revision{As one would expect}{In accordance with second law of thermodynamics (Sec.~\ref{sec:misconception})},
 there are no net spin currents flowing into the reservoirs in equilibrium, because the propagating spin-current is
 exactly compensated 
 by the (reflected) counter-propagating current.

 Upon applying a finite bias, states with energy inside the bias
 window are incoming only from one reservoir. In this case a spin
 current of one spin-conductance quantum (per angular momentum channel)
 survives inside the reservoirs. Also with respect to
 spin-conductances the \figref{fig:scattering}(a)
 and \figref{fig:coefficients} provide a consistent picture.

The qualitative discussion given here is fully backed up by an explicit calculation of all scattering probabilities;
for explicit results, we refer the reader to the \si{, Figure S2}.

%Although in principle the scattering probabilities do not contain direct information on the local observables inside the helix, the red and blue lines inside the helix (in the figure) are consistent with charge conservation at both interfaces and with the band-structure.
%Interestingly, it can be seen that in equilibrium there is no spin current in the leads because of the spin-flip reflections on both sides.
%
%Under small positive voltage, we can think of the transport as being induced by the incoming states from the left in \figref{fig:scattering}, while the incident electrons from right drop out. The spin currents in both leads can be easily recognized, consistently with \figref{fig:coefficients} for energies inside the gap. 
%
%It can be seen from the previous arguments that the spin-flip reflections play a crucial role. 

\subsubsection{Reservoirs accumulating spin\label{sec:acum}}
An implication of spin-flip scattering is that both reservoirs accumulate
spin in the presence of a current flow. To see how this happens we once again consult \figref{fig:scattering}(a). 
For a charge current flowing from $\Lld$ to $\Rld$, the drain acquires 
a spin up magnetization (red).
Simultaneously, 
the incoming flow of spin down
particles (blue) 
is reflected and spin flipped, so that the $\Lld$-lead  acquires a spin up magnetization (red), too. The observation represents a one-dimensional analogue of the Rashba-Edelstein
effect \cite{Edelstein1990}.

The parallel magnetizing of both leads is formally expressed by the anti-symmetry
of spin conductances
\begin{equation}
G_z^{(\Rld)} = - G_z^{(\Lld)}.
\end{equation}
The identity can be proven to hold under general conditions (\eg{}
at arbitrary energy) provided 
 the coupling to both leads is symmetric. We present a
formal proof in the \si{, Sec.~S5}.
Importantly, if mirror symmetry
is broken, the sign of conductances remains opposite, as we
demonstrate in the Sec.~\ref{sec:barrier}.

\subsubsection{Comparison with a spin filter}
For further illustration, we confront
the spin-flip scattering in the helical junction with the more familiar case of a magnetized junction that operates as a spin-filter, \figref{fig:scattering}(b). 
Also in this archetypal situation the reservoirs accumulate spin, however they do so in opposite directions (see the schematics of scattering in \figref{fig:scattering}). The key difference to the previous case is that spin is conserved everywhere, so that the (minority) spins 
%leaking 
removed from the source accumulate in the drain.

\subsection{\label{sec:barrier}Magnetizing reservoirs with imperfect contacts }
Realistic molecular junctions exhibit contact resistances that are not included in the minimal model.  %the transmission can drop orders of magnitude below unity due to spin-less reflection and the argument of the previous section is difficult to apply.
To investigate the effect of contact imperfections we add a potential term to the $\Lld$ contact,
\begin{align}
v(s') &= c\,\delta(s') \cdot \hat\sigma_0.
\label{e17}
\end{align}
This potential barrier is readily built into the scattering formalism as a modification of the matching conditions between the wire and the leads 
(\si{, Sec.~S3}).
\begin{figure}[t]
\centering
\includegraphics[width=1.0\columnwidth]{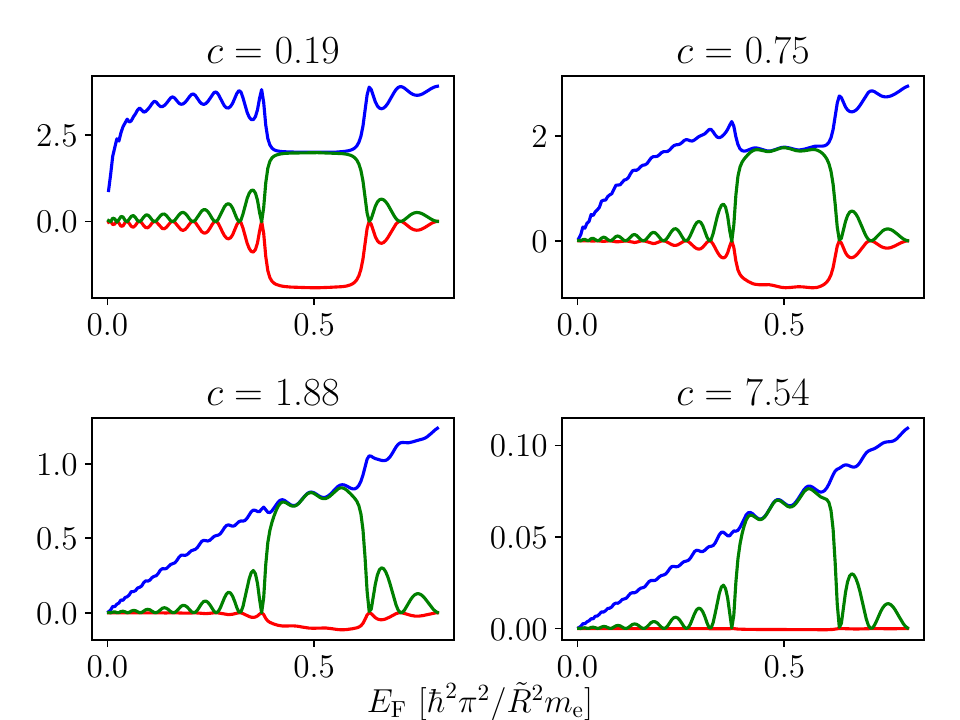}
\caption{\label{fig:opaque} 
Plot similar to Fig. \ref{fig:coefficients} illustrating the sensitivity of charge and spin conductances, $G/\frac{e^2}{h}$ (blue),  $G_z^{(\Rld/\Lld)}/\spinq$ (green/red) to the contact imperfection \eqref{e17} for a helical wire with 6 turns. 
The panels differ by increasing strength of the potential barrier $c$ at the left contact.
}
\end{figure}

\figref{fig:opaque} displays the conductances for increasing barrier strength. The charge conductance exhibits a gradual cross-over from the transparent, weak-barrier limit, $c\ll 1$, to the strong-barrier limit, $c\gg 1$, in which the transmission grows linearly with $\EF$.
Concomitantly, the spin-conductances evolve in a strikingly asymmetric fashion. 
For qualitative insight to the strong barrier limit we consult again Fig. \ref{fig:scattering}(a).
 For a current flowing from $\Lld$ to $\Rld$ and $\EF$ inside the spin-orbit gap, the barrier suppresses the transmitted current and spin-flip processes alike; hence, $G_z^{(\Lld)}$ is strongly suppressed inside the source. Conversely, the spin-current flowing in the drain equals the transmitted current (in units of the conductance quanta).  
Our argument implies that the spin conductances continue to exhibit
opposite signs in the presence of asymmetries. 
Therefore, we propose that the property of source and drain to magnetize
into the same direction upon a current flowing is a  general result
robust with respect to generic deformations of the minimal model
\eqref{e17}.

\subsection{Beyond the minimal model: Spin-flip reflection processes}
The qualitative analysis of scattering from the Fig.~\ref{fig:scattering}
indicates that spin-flipping reflections play a crucial role
in the spin transport within the minimal model. Indeed, within our phase coherent approach we can cast the 
spin conductances into the form 
\begin{align}
G_z^{(\Rld)} &= \spinq\sum_m
\qty(\Rr{\sd}{\su} - \Rr{\su}{\sd})\\
G_z^{(\Lld)} &= \spinq\sum_m
\qty(\Rl{\sd}{\su} - \Rl{\su}{\sd})
\end{align}
(see \si{, Sec.S5.3}). 
%This expression demonstrates that spin conductances and reflections are closely related to each other; one cannot exist without the other. 
\begin{comment}
%%% changed motivated by input by Bart von Wees
Note that these formul\ae{} are independent on the details of the junction's Hamiltonian
and only assume the presence of two orbital channels, labelled by $m$, related
by TRI. Therefore it is not surprising that the close connection between spin-currents and reflectivities
also holds in other models. For instance, 
Yang, van der Wal and van Wees have
demonstrated within an incoherent single channel model that spin-flip reflections must be non-zero for
spin-currents to exist in a chiral TRI conductor \cite{Yang2019}.
\end{comment}
Note that these formul\ae{} are independent on the details of the junction's Hamiltonian
and only assume the presence of two orbital channels, labelled by $m$, related
by TRI. Therefore it is not surprising that the close connection between spin-currents and reflectivities
also holds in other models. For instance, 
Yang, van der Wal and van Wees already gave arguments that spin-flip reflections must be non-zero for
spin-currents to exist in a chiral TRI conductor \cite{Yang2019}.
Our work goes significantly beyond, since we prove that the spin-conductance per channel is essentially identical with the spin-flip  reflectance.

\subsection{Induced spin and charge densities in the contact}
We supplement the transport results with expressions
for the charge and spin density in the $\Lld$-lead $(s<0)$ in linear response. The latter quantities
will be given in units of $e$ and $\hbar/2$, respectively.
Our formul\ae{} follow from scattering theory and are thus independent
on the microscopic details of the Hamiltonian.
The details of the derivation are moved to
\si{, Sec.~S6}, for the sake of brevity.

The charge density evaluates to the expression
\begin{equation}
n(s) = n(s)\Bigr|_{V=0}
+ V\, \varrho_\mathrm L(\EF) \Biggl[\qty(4 - \frac{h}{e^2}G)
+
2\sum_m \Re{e^{-\iu 2qs} r_{\su\su,m}^{(\Lld\Lld)}} \Biggr]_{E=\EF} 
\ {+}\,  \mathcal O\qty( V^2 ).
\label{eq:n0}
\end{equation}
where $r_{\su\su,m}^{(\Lld\Lld)}=r_{\sd\sd,-m}^{(\Lld\Lld)}$ has been used reflecing TRI
(see also \si{, Sec.~S6}).
The first term is the equilibrium charge density; 
as shown in \si{, Sec.~S6.2}, it reveals a
familiar contact Friedel oscillation 
%(power-law decay $\propto 1/s$)
caused by quantum interference
with reflected waves in a Fermi ground-state.

In the linear response term in \pref{eq:n0} we factored out the expression
\begin{equation}
\label{eq:ldosl}
\varrho_\mathrm L(\EF) = \frac 1{2\pi}
\sum_{q>0}\delta
\qty(\EF - \frac{q^2}2) =
\frac 1 {4\pi^2\qF},
\end{equation}
being a (homogeneous)
local density of states of right movers (per spin) in a single channel homogeneous wire.
The square brackets in \epref{eq:n0} contain two terms: a homogeneous ($s$-independent) term, $\qty(4 - \frac{h}{e^2}G)$, represents the enhancement of the charge-density that is associated with the reflected particles. The reflection-induced enhancement of the charge density also leaves a trace in the Friedel-oscillations, which is expressed by the second term.  

Notice that unlike spin-conserving processes, spin-flip processeses, such as incorporated by 
$r_{\sd\su,m}^{(\Lld\Lld)}$, do not contribute 
to the oscillating term in \eqref{eq:n0} because the superposition of two probability currents with opposite spins has no cross (interference) term. Spin-flip terms do enter the conductance, $G$, of course. 
 
The induced spin density in the left lead reads
\begin{equation}
\label{eq:dnz}
n_z(s) = -V\, \varrho_\mathrm L(\EF) \frac{4\pi}e G_z^{(\Lld)}
\, +\, \mathcal O\qty(V^2); 
\end{equation}
there is no spin density in equilibrium because of TRI.
%This statement is an expression of the fact that the Fermi ground state of the junction is non-degenerate.
The (linear) spin imbalance represents the the loss of spin-density associated with the transmitted spin-current. For this reason, Eq. \eqref{eq:dnz} implies that a measurement of the linear response of the local spin-density yields (up to trivial factors) the transport coefficient $G_z^{(\Lld)}$. 
Notice that in striking contrast to the charge density, the spin density does not display Friedel oscillations even in linear response; ultimately, as we show in the \si{}, the reason is TRI. 
%
%We further mention that for a helix with $\helc = -1$ and $V>0$, the induced spin density is ``up'' and at the same time $I^{(\Lld)} <0$. Hence, these observables offer a complementary picture of the $\Lld$-lead.

We further note that the current-induced spin-accumulation, \eqref{eq:dnz}, goes hand in hand with a spin-split chemical potential if a self-consistent description of the quasi-static non-equilibrium situation is employed \cite{Nazarov2009}: 
\begin{equation}
 \mu_\su^{(\Lld)}(V) - \mu_\sd^{(\Lld)}(V)
\ {=}\ -V \frac{\pi}{2e}G_z^{(\Lld)}\, +\, 
\mathcal O \qty(V^2); 
\end{equation}
the formula follows from \eqref{eq:dnz} and is also derived in the \si{, Sec.~S6}. 
We offer two comments on this expression:
First, the coefficient multiplying $V$ can reach $\frac 14$
if $\EF$ lies in the SOI gap, \ie{} the spin accumulation is 
non-analytic in the SOI strength. 
Second, assuming that the
reservoirs ultimately relax spin (and charge), but do not introduce
backscattering, the spin 
accumulations can be measured in the vicinity
of the contact by a four-terminal setup, or by Hanle spin precession,
as suggested by Yang, van der Wal and van Wees\cite{Yang2020,Yang2021}.

%%%%%%%%%%%%%%%%%%%%%%%%%%%%%%%%%%%%%%%%%%%%%
\section{Further remarks}

\begin{itemize}[leftmargin=0.5cm,topsep=0em]
\setlength\itemsep{-0.0em}
\item 
\revision{The (experimental) measurementof spin}{Spin} currents and spin accumulation \revision{is}{are} a
central topic in the field of {\it Spintronics}.
\cite{Bardarson2007,Adagideli2009} In particular, the possibility of spin
accumulation near interfaces between materials with and without spin-orbit
interaction is well understood \cite{Adagideli2007}; the results we report here
on the minimal level confirm the validity of the general picture down to the
molecular scale. \revision{}{With an eye on experiments, we mention that the amount of accumulated spin should depend on the spin-relaxation time. Our result for the spin accumulation applies to local measurements a distance no longer than the spin relaxation length away from the contact.}

In Sec.~2 we have mentioned that the bands for
energies in the spin-orbit gap are analogous to edge states associated with a
quantum spin Hall device. As a consequence, the minimal model exhibits
spin-momentum locking with the consequence that backscattering off defects is
suppressed. In other words, we expect that our transport results are robust
against weak (non-magnetic) disorder.

\item
Recently, a related work 
on parallel 
 spin-accumulation
%magnetizing 
and spin transport
has been published \cite{Wolf2023}. The tight-binding chain investigated by these authors is essentially equivalent to our toy model with a distinctive difference: in Ref.~\cite{Wolf2023} the SOI is confined to the contact bond, only, while in our toy model the SOI is a property of the molecule and non-vanishing along the entire helical structure. 

The  spin conductances reported in Ref.~\cite{Wolf2023} show the same symmetry, $G_z^{(\Lld)}=-G_z^{\Rld)}$, as in our work.
%as expected based on reasons of TRI. 
However, the energy dependences of the spin conductances exhibit pronounced differences:  $G_z^{(\Lld/\Rld)}(E)$ in Ref.~\cite{Wolf2023} does not exhibit a fixed sign; in our case it does, so that the sign of the spin accumulation
is independent of the Fermi energy.
This difference can be traced back to the different ways how SOI is implemented; it may, at least in principle, be used to discriminate one situation from the other, experimentaly. 

\item 
It is well-known in Spintronics that spin conductances are non-zero in
junctions with multiple conduction channels only (see \si{, Sec.~S1}, Principle
1). In our model, the absence of spin polarization for a single channel is
immediately obvious by setting $m=0$ in Eq.~\eqref{e7}.
Molecular junctions generically exhibit multiple channels and are thus
prepared for hosting spin currents (\eg{} see Ref.~\cite{Liu2021} for a channel
analysis of helical peptides).
Frequently employed molecular linker units
(\eg{} thiol groups linked to Au) effectively suppress all conduction channels
but one. Moreover, the generation of significant spin currents benefits from
two conduction channels of similar transmission probabilities, which in turn
requires quasi-degenerate molecular orbitals. 
To favor such conditions, linker-free benzene-type structures are natural
candidates. They can be functionalized with heavier elements (see \eg{}
Ref.~\cite{Adachi2017}) to boost the spin-orbit coupling and promote chirality.
\end{itemize}

%%%%%%%%%%%%%%%%%%%%%%%%%%%%%%%%%%%%%%%%%%%%%
\section{Summary and conclusions}
We have investigated charge and spin transport in a minimal model of a helical
molecule with spin-orbit coupling attached to two spin-conserving leads. The
minimal model was first devised by Michaeli and Naaman and allows for a full
analytic treatment of transport properties. While the earlier authors have
focussed on spin polarizations, we calculated the full conductance matrix,
including charge and spin conductances. 

The band-structure of the minimal model hosts four helical bands that exhibit
spin-momentum locking. We first clarify the connection between transport
properties and band-structure.  In particular, first glances could suggest the
existence of an equilibrium spin current, which sometimes is used as an
argument against the validity of the model. We explain the origin of the
misunderstanding, which results from neglecting the contact scattering that
always exists in a transport geometry. In actuality, there is no  equilibrium
spin transport in this model. 

Our explicit calculation of charge and spin conductances in the Landauer
formalism show that the spin conductance reaches a maximum ($e/2\pi$) for
energies inside the gap, corresponding to 2 fully polarized conduction
channels. Outside the gap region the spin conductances remain sizable, too.
Further, we find that at small biases and for incoming currents being
unpolarized, there are spin currents with opposite signs in each lead. 

Upon a current flowing, spin
accumulates in the vicinity of each contact as a consequence of the spin
currents; the magnetizations at left and right contacts are the same and
reverse with voltage and helicity. The magnitude of the accumulated
spin polarization is directly proportional to the spin conductance
of the respective lead; therefore, we reveal a new route towards
 the measurement of $G_z^{(\Lld/\Rld)}$, 
as has also been proposed for specific device geometries by Yang {\it et al.} \cite{Yang2021}.

\begin{comment}
We add a remark on spin-accumulation in mesoscopic systems: accumulated spins
have a feedback on the transport currents in first order in the applied biases.
In particular, spin accumulations can drive charge currents as well as charge
currents can drive spin currents, reflecting an Onsager reciprocity.
Seemingly, this reciprocity has not been widely appreciated in
previous theories of the CISS effect.
\end{comment}

The minimal model therefore directly reveals important features of analytic
structure of spin transport in helical molecules. It can serve as a
guidance for the interpretation of ab-initio calculations of transport
coefficients in chiral molecular systems, \eg{}, based on
nonequilibrium Green's functions. Further, it lends
itself to straightforward generalizations, e.g.,  the analytic
calculation of current-induced mechanical torques in a quantum model. 

%%%%%%%%%%%%%%%%%%%%%%%%%%%%%%%%%%%%%%%%
\begin{comment}
\section{Methods}
The band-structure of an infinite helical wire was obtained by 
diagonalizing the \epref{e8}. Explicit analytic expressions
are given in the \si{, Sec.~{S2}}. 
The junction was modelled with a central helical element of length $L$
attached to a pair of semi-infinite straight tubes (\figref{fig:hcoord}).
Hamiltonian 
eigenstates were obtained by wavefunction matching, see Sec.~{S3}.
Spin conductances \pref{eq:coeffs} were obtained from the scattering matrix
in Landauer approach; the respective formul\ae{} were re-derived
in Sec.~S5 for reference purposes. The local magnetization density
was calculated from explicit wave-functions in presence of
two chemical potentials, Sec.~S6.
\end{comment}

\section*{Supporting Material} The accompanying file contains
supporting sections S1-S6 with methodological details.

\begin{acknowledgments}
Financial support for the project was provided by the the Czech Science
Foundation (project no. 22-22419S), the Netherlands Organisation for Scientific
Research (NWO), grant 680.92.18.01. FE acknowleges support from the German
Research Foundation (DFG) through the Collaborative Research Center, SFB 1277
(project A03), through GRK 2905, project-ID 502572516
 and through the State Major Instrumentation Programme, INST
89/560-1; project number 464531296. 
Discussions with K. Michaeli, O. Tal, K. Richter, J. Fabian, J. Schliemann, D. Weiss, W. Wulfhekel, 
D. Kochan,
B. Yan are acknowledged.
We thank I. Dimitriev for correcting our expression of the current
operator.
\end{acknowledgments}

\bibliography{references}

\begin{thebibliography}{33}
\bibitem [{Notei()}]{Notei}%
  \BibitemOpen
  \bibinfo {note} {As is well known, due to the Onsager relations
  magnetotransport experiments based on the polarizer-analyzer setup will not
  detect spin-currents in the linear transport regime: even if a non-vanishing
  spin-current is flowing, the charge conductance is independent of the
  magnetization direction $G(\protect \mathbf {M})=G(-\protect \mathbf {M})$
  and therefore insensitive to the flow of spin.}\BibitemShut {Stop}%
\bibitem [{Noteii()}]{Noteii}%
  \BibitemOpen
  \bibinfo {note} {The details of the confinement $V_\protect \mathrm H(\varrho
  )$ are not relevant for the main conclusions of our work. Focusing on the
  lowest-energy eigenstates, we adopt a harmonic approximation $V_\protect
  \mathrm H(\varrho ) \approx \protect \frac 12\protect \frac {\hbar
  ^2}{m_{\protect \mathrm e}d^4}\varrho ^2$: the resulting wavefunctions $\Phi
  _{N,m}(\theta ,\varrho )$ are eigenstates of a 2D harmonic oscillator with
  $m$-independent eigen-energies $E_N$ and $m$ restricted to $ -N, -N+2,
  \protect \ldots , N-2, N$.}\BibitemShut {Stop}%
\bibitem [{Noteiii()}]{Noteiii}%
  \BibitemOpen
  \bibinfo {note} {All operators transform as \begin {align*} \protect \hat A
  \DOTSB \mapstochar \rightarrow e^{\protect \mathrm i{\lambda } \protect \hat
  \sigma _{z} s\pi /\protect \tilde R} \protect \hat A e^{-\protect \mathrm
  i\lambda \protect \hat \sigma _{z} s\pi /\protect \tilde R}. \end {align*} To
  understand the action of the exponential operator on the spin raising and
  lowering operators in Eq. \protect \eqref {e5}, one recalls that $\protect
  \hat \sigma _{+}$ ($\protect \hat \sigma _{-}$) annihilates spin up (spin
  down) states. Therefore the transformed operator \begin {align*} e^{\protect
  \mathrm i{\lambda } \protect \hat \sigma _{z} s \pi /\protect \tilde R}
  \protect \hat \sigma _{+} e^{-\protect \mathrm i\lambda \protect \hat \sigma
  _{z} s\pi /\protect \tilde R} \end {align*} annihilates a spin-up state - as
  it would without being transformed - or equip a spin-down state with two
  times the same phase factor, i.e. $e^{2\protect \mathrm i\lambda s\pi
  /\protect \tilde R}$. As a consequence, the transformation indeed removes the
  phase factor of $\protect \hat \sigma _{+}$ seen in Eq. \protect \eqref
  {e5}.}\BibitemShut {Stop}%
\end{thebibliography}
\begin{thebibliography}{10}
\bibitem [{Noteii()}]{Noteii}%
  \BibitemOpen
  \bibinfo {note} {We remind that the operator $\protect \hat \sigma _z =
  \protect \mathrm i\exp (-\protect \mathrm i\protect \hat \sigma _z\pi /2)$
  performs a rotation by $\pi $ around the $z$ axis of the spin degrees of
  freedom (up to a phase factor); it effectively inverts the sign of $\protect
  \hat \sigma _y$. Such rotation can be compensated for by flipping the sign of
  $m$.}\BibitemShut {Stop}%
\end{thebibliography}

\end{document}

\author{Richard Korytár}%
\email{richard.korytar@ur.de}
\affiliation{Department of Condensed Matter Physics, Faculty of Mathematics and Physics, Charles University, Ke Karlovu 5, 121 16, Praha 2, Czech Republic}

\author{Jan M. van Ruitenbeek}
\affiliation{Huygens-Kamerlingh Onnes Laboratory, Leiden University, NL-2333CA Leiden, Netherlands}

\author{Ferdinand Evers}
\affiliation{Institute of Theoretical Physics, University of Regensburg, D-93050 Regensburg, Germany}

\title{Spin conductances and magnetization production in chiral molecular junctions}

%\keywords{Chirality-induced spin selectivity, molecular junctions, spintronics}
\maketitle

\section{Review of symmetry properties of transport coefficients in
 molecular junctions}

Principles of charge and spin transport in two-terminal devices
were derived in the fields of nonequilibrium statistical mechanics
and spintronics. Here we recollect some
of them that are directly relevant for helical molecular
junctions with SOC. \textbf{Principles 1} and \textbf{2}
 impose tight restrictions on the generation and detection of spin currents in nanoscale conductors.
 In \textbf{Principle 3} we show
 that TRI implies reciprocity relations between certain spin transport
 coefficients.

\paragraph{Principle 1: Absence of spin currents and spin filtering in single-channel
non-magnetic junctions.}
We consider a strictly one-dimensional conductor with spin-orbit coupling.
Assuming a single-particle scattering scenario, it can be shown that
an unpolarized incident electron flux can not lead to polarized outgoing
fluxes if the leads carry a single channel. 
This was shown by Kiselev and Kim \cite{Kiselev2005} as a consequence of
time-reversal invariance.
Therefore, spin currents can not be induced by the voltage bias
even in the non-linear regime, unless the above conditions are relaxed.
The same negative statement applies to spin filtering.

\paragraph{Principle 2: Absence of chirality-induced linear magnetoconductance.}
Now let us consider attaching a ferromagnetic lead with magnetization $M$
to the helix. 
%The analyzer transmits majority-spin electrons 
%and reflects minority-spin. 
The number of conduction channels can be
arbitrary. Of interest is the difference between currents
at reversed magnetizations: $I(M) - I(-M)$ (magnetocurrent) and
%a respective 
the related 
difference in linear conductances, $G(M) - G(-M)$ (magnetoconductance).

A system with magnetization $M$ and a current flowing has the property
that the modulus of the current is invariant under time reversal, \ie{} if
one reverses the velocities of all particles and inverts the sign of $M$.
A consequence of this invariance is the well-known Onsager's relation for the charge conductance\cite{Onsager31, Casimir1945,Nazarov2009},
\begin{equation}
\label{eq:onsager}
G(M) = G(-M).
\end{equation}
Note that the relation does not rely on the assumption of phase-coherence, unlike \textbf{Principle 1}.
Furthermore, the relation also holds in presence of a magnetic field, if the
latter is reversed along with $M$.
It was recognized by Yang \etal \cite{Yang2019} that \epref{eq:onsager} prohibits the
observation of spin filtering of chiral molecules in the linear
response in the above mentioned setup.
Nevertheless, magnetocurrent can be non-zero in the non-linear regime.

Onsager's theorem also precludes magnetoconductance in standard
\textit{ab-initio} transport calculations even in the non-linear response
unless charge self-consistency with respect to the bias voltage is achieved.
We detail this argument in Section~\ref{sec:abinitio}.

\paragraph{Principle 3: Reciprocity of spin-transport coefficients.}
Linear response relations shown in the main text involve charge and spin currents
reacting to external voltage bias.
In order to introduce reciprocity we need a slightly more general formulation
that includes spin dependent chemical potentials.
Following Nazarov and Blanter \cite{Nazarov2009},
we introduce \textit{spin accumulations}
as differences in the respective chemical potentials between each spin species (in the same lead).
Namely, the $z$-components of the spin accumulations are
\begin{subequations}
\begin{align}
\saW^{(\Rld)} = \qty(\mu_\su^{(\Rld)} - \mu_\sd^{(\Rld)})/\hbar,\\
\saW^{(\Lld)} = \qty(\mu_\su^{(\Lld)} - \mu_\sd^{(\Lld)})/\hbar.
\end{align}
\label{eq:imbalances}
\end{subequations}
Linear response of the currents to the voltage and spin accumulations
is characterized by coefficients of the expansion
\begin{comment}
\begin{alignat}{4}
I(V, \saW^{(\Rld)}, \saW^{(\Lld)})\ &{=}\ GV\ &&{+}\ \widetilde G_z^{(\Rld)}\saW^{(\Rld)} 
\ &&{+}\ \widetilde G_z^{(\Lld)}\saW^{(\Lld)} &&{+} \ldots\\
I_z^{(\Lld)}(V, \saW^{(\Rld)}, \saW^{(\Lld)})\ &{=}\ G_z^{(\Lld)}V\ &&{+}\ \coefzz^{(\Lld\Lld)}\saW^{(\Lld)}
\ &&{+}\ \coefzz^{(\Lld\Rld)}\saW^{(\Rld)} &&{+} \ldots\\
I_z^{(\Rld)}(V, \saW^{(\Rld)}, \saW^{(\Lld)})\ &{=}\ G_z^{(\Rld)}V\ &&{+}\ \coefzz^{(\Rld\Lld)}\saW^{(\Lld)}
\ &&{+}\ \coefzz^{(\Rld\Rld)}\saW^{(\Rld)} &&{+} \ldots,
\end{alignat}
\label{eq:linearresponse}
\end{comment}
\begin{equation}
\begin{array}{rcrcrcrl}
I(V, \saW^{(\Rld)}, \saW^{(\Lld)})\ & {=}\ & GV\ & {+}\ & \widetilde G_z^{(\Lld)}\saW^{(\Lld)}\ & {+}\ & \widetilde G_z^{(\Rld)}\saW^{(\Rld)} 
\ & {+}\ \ldots\\
I_z^{(\Lld)}(V, \saW^{(\Rld)}, \saW^{(\Lld)})\ & {=}\ & G_z^{(\Lld)}V\ & {+}\ & \coefzz^{(\Lld\Lld)}\saW^{(\Lld)}
\ & {+}\ & \coefzz^{(\Lld\Rld)}\saW^{(\Rld)}\ & {+}\ \ldots\\
I_z^{(\Rld)}(V, \saW^{(\Rld)}, \saW^{(\Lld)})\ & {=}\ & \ G_z^{(\Rld)}V\ & {+}\ & \coefzz^{(\Rld\Lld)}\saW^{(\Lld)}
\ & {+}\ & \coefzz^{(\Rld\Rld)}\saW^{(\Rld)}\ & {+}\ \ldots,
\end{array}
\label{eq:linearresponse}
\end{equation}
where we suppressed terms of higher order. Naturally, additional measurements of spin in other directions
than $z$ would bring up coefficients in these directions, too, but they are of no relevance for the present work.
See Ref.~\cite{Jacquod2012} for a general formulation. Time-reversal invariance implies
the reciprocities\cite{Jacquod2012}
\begin{subequations}
\begin{align}
\label{eq:reciprocity1}
G_z^{(\Rld)} &= +\widetilde G_z^{(\Rld)}\\
\label{eq:reciprocity2}
G_z^{(\Lld)} &= -\widetilde G_z^{(\Lld)}\\
\label{eq:reciprocity3}
\coefzz^{(\Lld\Rld)} &= - \coefzz^{(\Rld\Lld)}.
\end{align}
\label{eq:reciprocity}
\end{subequations}
We remark
that the difference in signs of the right-hand sides above results
from our definition of the currents that measure
particles entering the junction through the left contact, or,
leaving the junction through the right contact.

In molecular junctions, these relations offer an alternative
 way to measure the spin currents in linear response.
Namely, the $\Lld$ and $\Rld$ spin currents induced by a finite $V$ but zero $\saW^{(\Rld/\Lld)}$
are identical, up to a sign, to
the charge current that results from the spin accumulation
in the $\Lld$ and $\Rld$ leads, respectively, at $V=0$.
The reciprocity relations have been applied to the CISS transport by Yang, Wal and Wees in Ref.~\cite{Yang2020}.
Further reciprocities emerge with the measurement of heat currents
and temperature differences; see
Ref.~\cite{Jacquod2012} for a review.

\par We re-derive explicit expressions for the transport coefficients in
Landauer formalism in
Section~\ref{sec:landauer} of this \si{}. In Section~\ref{sec:symmetries}
we review various symmetry properties of the scattering operator
that apply to the Michaeli-Naaman model and employ them to derive
cross-relations between spin conductances in Sec.~\ref{sec:landauer}, including
Onsager relations.
Explicit evaluation of the scattering operator for the model is detailed in 
Sections~\ref{sec:diag}, \ref{sec:eval}.

\subsection{\label{sec:abinitio}Absence of chirality-induced magnetocurrent in common
\textit{ab-initio} transport calculations}
The state-of-the-art \textit{ab-initio} method employs the Landauer formalism
extended to non-equilibrium. It assumes quantum coherence,
effective single-particle states
and thermodynamic equilibrium in each lead.
The electric current is given by the formula
\begin{equation}
\label{eq:current}
I(V,M) = \frac eh \int \Bigl[ \nf{\mu_\Lld(V)} - \nf{\mu_\Rld(V)}\Bigr]
\sum_{\sigma\sigma'}T_{\sigma\sigma'}(E,V,M)\dd{E} .
\end{equation}
The central quantity in the above equation is the spin-resolved
transmission probability $T_{\sigma\sigma'}(E,V,M)$. We explicitly
indicate its dependence on the voltage $V$ and a static magnetization of
the junction $M$.\footnote{%
The scattering probabilities are further differentiated according to
the direction of flow, \ie{} from left to right or \textit{vice versa}; see the analysis
of the scattering matrix of the helical model in the \si{}.
Both directions of flow yield the same \textit{charge current}
and therefore we suppress the direction of flow in the notation of
$T_{\sigma\sigma'}$.
}
The $T_{\sigma\sigma'}(E,V,M)$ results from a quantum-mechanical
calculation, \eg, by the non-equilibrium Green's function technique.
The remaining factor in the integrand contains the voltage-dependent
energy window allowed by the Pauli principle. The Fermi-Dirac
distribution $n_\mathrm F$ depends explicitly on the chemical potential
of each lead, $\mu_{\Lld/\Rld}(V)$. The boundary condition $\mu_\Lld(V)
-\mu_\Rld(V) = eV$ is imposed by the external bias voltage $V$.
In this discussion we suppress the temperature dependence and assume
that both leads have equal temperatures.

The \textit{ab-initio} calculations based on density-functional theory
deliver both $\mu_{\Rld/\Lld}(V)$ and $T_{\sigma\sigma'}(E,V,M)$ from the
Kohn-Sham system.
An explicit calculation of the voltage dependencies in the previous formul\ae{}
is rarely performed. A frequently employed approximation
\begin{equation}
\label{eq:nostark}
T_{\sigma\sigma'}(E,V,M) \approx  T_{\sigma\sigma'}(E,0,M)
%,\qquad %\mu_{\Rld,\Lld}(V) = \mu \pm \half eV
\end{equation}
when plugged into \epref{eq:current}, still delivers a non-linear
$IV$ curve.
%However, $I(V,M) = - I(-V, M)$, \ie this approximation does not
%describe rectification.

Time-reversal invariance implies the following symmetry
property of the transmission probability\cite{Jacquod2012}
\begin{equation}
\sum_{\sigma\sigma'} T_{\sigma\sigma'}(E,0,M) = \sum_{\sigma\sigma'} T_{\sigma\sigma'}(E,0,-M).
\end{equation}
It follows trivially that the magnetocurrent $I(V,M) - I(V,-M)$ vanishes 
when the approximation
\pref{eq:nostark} is used. We conclude that the common \textit{ab-initio}
methodology can not describe chirality-induced nonlinear magnetocurrent
due to Onsager's theorem.

%%%%%%%%%%%%%%%%%%%%%%%%%%%%%%%%%%%%%%%%%%%%%%%%%%%%%%%%%%%%%%%%%%%

\section{\label{sec:diag}Spectrum and eigenstates of a helical tube}
The Hamiltonian, Eq.~(\eqhamk) of the main text,
can be rewritten as
\begin{comment}
\begin{equation}
    \hat H(k) = \left(\half + \half k^2\right)\,\hat\sigma_0 + \helc\begin{pmatrix}
    - k & -\iu\tilde\kappa m \\
    \iu \tilde\kappa m & + k
    \end{pmatrix}
\end{equation}
\end{comment}
\begin{equation}
    \hat h_m(k) = \left(\half + \half k^2\right)\,\hat\sigma_0 + \helc {\mathfrak b}\cdot{\boldsymbol{\sigma}}
\end{equation}
where the first term on the right side is proportional to the unit matrix $\hat\sigma_0$; $\boldsymbol{\sigma}$ abbreviates the Pauli-matrices and ${\mathfrak b}=(b_x,b_y,b_z)=(0,-\tilde\kappa m,-k)$.

The second term has eigenvalues
\[
 \helc k\xi_\alpha,\quad \text{where }\xi_\alpha = \alpha \sqrt{ 1 + \left(\frac{\tilde\kappa m}{k}\right)^2}
 \text{ for } \alpha = \pm 1.
\]
The eigenvectors read
\begin{equation}
\label{eq:vec}
\Psi_{m\alpha} (k) = \frac \helc
{\sqrt{k^2 + \left(\frac {\tilde\kappa m}{1 + \xi_\alpha}\right)^2}}
\begin{pmatrix}
\iu \frac {\tilde\kappa m}{(1 + \xi_\alpha)} \\
k
\end{pmatrix}.
\end{equation}

We chose the sign of the eigenvalue, $\alpha\helc\cdot\sgn{k}$, and the phase of
the eigenspinors such that
for $k \rightarrow \pm \infty$ or $\tilde\kappa \rightarrow 0$ the
eigenstates approach up, down spinors, $\alpha =\sigma$.
The eigenenergies of the Hamiltonian  are
\begin{align}
\label{eq:disp1}
E_{m,\alpha}(k) &= \half\left(k + \helc\xi_\alpha\right)^2 -
\half
\left(\frac{\tilde\kappa m }{k}\right)^2
\end{align}
For small $\tilde\kappa$, it resembles the spectrum of the straight tube, except that
there is a gap $E \in (\half - |\tilde\kappa m |, \half +| \tilde\kappa m |)$ around $k=0$.
The inverse dispersion reads
\begin{align}
\label{eq:inverse2}
k_{m\alpha\beta} &=  \zeta\helc \alpha \sqrt{2\left[ E +\half +\beta\sqrt{2E +\tilde\kappa^2 m^2 }\right]}\\
\zeta &= \left[ \sgn{E - \half}\right]^{\frac{(1-\beta)}{2}}
\end{align}
and it is labeled by the branch $\beta$ for each eigenstate $\alpha$.

\begin{figure}
\includegraphics[width=0.5\columnwidth]{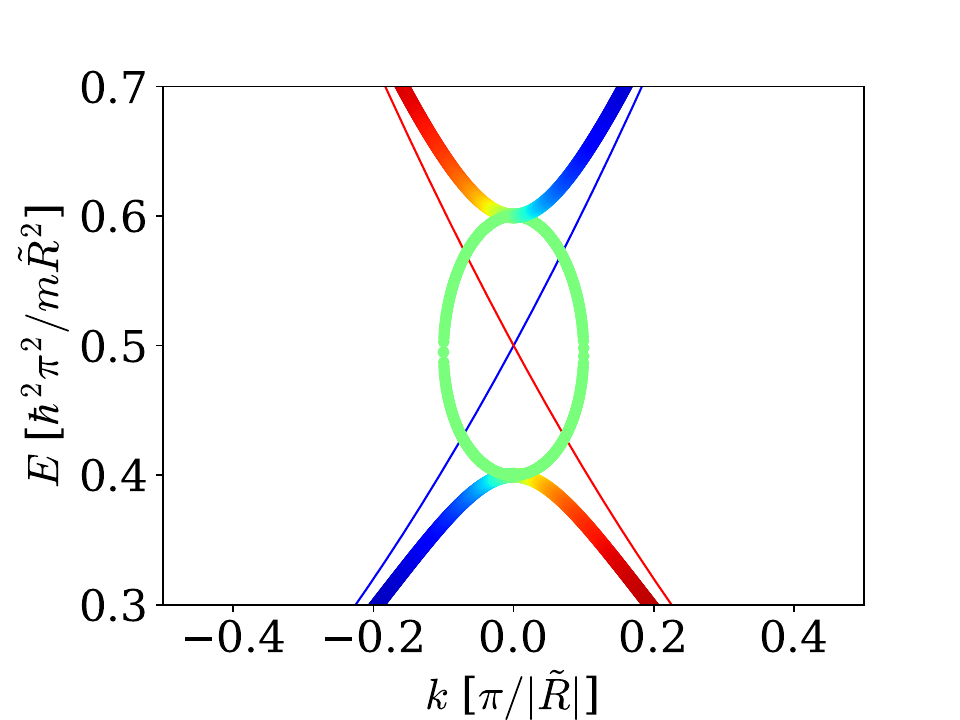}
\caption{\label{fig:eva}Dispersion of the helix in the region of the
spin-orbit gap. The parameters and the symbols are identical to the
Fig.~3 of the main text. Additionally, we plot the energies of the 
evanescent states within the gap (smaller dots). }
\end{figure}

Real energies of evanescent modes and their eigenspinors can be obtained
by analytic continuation of the expressions into the complex $k$-domain.
The dispersion of the evanescent modes is shown in \figref{fig:eva}.
The expectation value of $\hat\sigma_z$ is zero in these states.
An inspection shows that their spin vector rotates in the $xy$-plane.

\section{\label{sec:eval}Evaluation of the scattering matrix of the helical junction}
Expressions for the scattering wavefunctions are derived in this section
by matching the current operator on both sides of the interfaces.
The wave function coefficients deliver scattering matrix elements
straightforwardly. Throughout the following sections  we employ dimensionless energies
and lengths introduced in the main text but drop the primes from the
notation.
\paragraph{Probability current.}
In the gauged Hilbert's space the time evolution is generated by
the Hamiltonian Eq.~(\eqhams) of the main text. From the continuity equation
of a probability density it is possible to derive an explicit form
of the probability current
\begin{equation}
\label{eq:curg}
j(s,t) = \half\left\{
\Psi^\dagger\left(-\iu \partial_s + \tilde\gamma m - \helc\hat\sigma_z\right)\Psi
+
\left[\left(-\iu\partial_s +\tilde\gamma m - \helc\hat\sigma_z\right)\Psi\right]^\dagger
\Psi
\right\}
\end{equation}

The form of the current operator is analogous to
the one in the problem of an electron in a magnetic field, except that the gauge potential
is matrix-valued (non-Abelian).

The above expression for the current operates on the gauged wavefunctions in the helix.
Upon employing the inverse gauge transformation $\Psi' = e^{-\iu s\helc\hat\sigma_z}\Psi$
in \epref{eq:curg}, we obtain an expression
\begin{equation}
\label{eq:curu}
j'(s,t) = \half\left\{
\Psi^{'\dagger}\left(-\iu \partial_s + \tilde\gamma m \right)\Psi'
+
\left[\left(-\iu\partial_s +\tilde\gamma m \right)\Psi'\right]^\dagger
\Psi'
\right\}
\end{equation}
where primes indicate ungauged expressions.
In the straight tubes, the current is given by the above expression with
$\tilde\gamma = 0$.

Furthermore, the Abelian gauge potential $\tilde\gamma m$ can be removed from the
Hamiltonian by the gauge transformation $e^{\iu s \tilde\gamma m}$ in the Hilber's
space of the molecule.
Correspondingly, $\tilde\gamma$ will vanish from the current operator.
As a corollary, the expectation values
of the (spin) current density in Hamiltonian eigenstates are $\tilde\gamma$ independent.
Hence, when leads are attached, scattering probabilities will be likewise
$\tilde\gamma$ independent, although the scattering phases won't.
It is known that
in one dimension, a vector potential does not have physically observable
effects.
Therefore we set $\tilde\gamma = 0$ in what follows.

\paragraph{Scattering wavefunctions.}
The scattering picture is most transparent in the ungauged Hilbert's space.
There, the wavefunctions in the leads are incoming and outgoing planewaves with wavenumbers
$\pm \sqrt{2E}$. In the helix, we apply the inverse gauge transformation
$e^{-\iu\helc \hat\sigma_z s}$ to the eigenvectors, \epref{eq:vec}.

We consider a helix of length $L$ connected to straight semi-infinite tubes.
First, a spin-up electron is coming from the left tube with energy $E>0$.
In the left tube the wavefunction reads
\begin{equation}
\psi_m^E(s) =
\begin{pmatrix}
1 \\
 0
\end{pmatrix} e^{\iu qs} \ +\
\begin{pmatrix}
r_{\uparrow\uparrow}  \\
r_{\downarrow\uparrow}
\end{pmatrix} e^{-\iu qs},\quad s<0,\quad q=\sqrt{2E}.
\end{equation}
The first term represents an incident electron and
the second term represents a reflected electron, with reflection
amplitudes $r_{\sigma\sigma'}$, where the left (right) index labels the final (initial) state.

In the right lead ($s>L$), the transmitted wavefunction reads
\begin{equation}
\psi_m^E(s) =
\begin{pmatrix}t_{\uparrow\uparrow} \\
t_{\downarrow\uparrow}
\end{pmatrix} e^{\iu qs}.
\end{equation}
Similarly, for an incoming down electron the following relations hold
\begin{align}
\psi_m^E(s) &=
\begin{pmatrix}
 0\\
 1
\end{pmatrix}\, e^{\iu qs}\  {+}\
\begin{pmatrix}
r_{\uparrow\downarrow}  \\
r_{\downarrow\downarrow}
\end{pmatrix} e^{-\iu qs},\quad s<0\\
\psi_m^E(s) &=
\begin{pmatrix}t_{\uparrow\downarrow}   \\
t_{\downarrow\downarrow}
\end{pmatrix} \, e^{\iu qs},\quad s>L.
\end{align}
In the central (helical) region, the form of the wavefunction depends on whether we are in the
gap region or away from it. It consists always from four eigenstates, with
amplitudes $a,b,c,d$. Away from the gap, the wavefunction contains
planewaves only,
\begin{align}
\psi_m^E (s) &=\,
      a\, e^{\iu(k_{m++} -\helc\hat \sigma_z)s}\cdot \Psi_{m,+}(k_{m++})\\
 &+\, b\, e^{\iu(k_{m-+} -\helc\hat \sigma_z)s}\cdot \Psi_{m,-}(k_{m-+})\\
 &+\, c\, e^{\iu(k_{m+-} -\helc\hat \sigma_z)s}\cdot \Psi_{m,+}(k_{m+-})\\
 &+\, d\, e^{\iu(k_{m--} -\helc\hat \sigma_z)s}\cdot \Psi_{m,-}(k_{m--}),\quad 0<s<L
\end{align}
where the wavenumber corresponding to $E$ is given in the \epref{eq:inverse2}.
In the gap region, two modes become evanescent and the above expression
has to be analytically continued.
Because of the inverse gauge transformation operator, each component of the spinors
has a distinct spatial dependence.

The coefficients $a,b,c,d$ (for a given incoming spin)
and the transmission and reflection amplitudes
are determined by matching the expectation values of the probability current
at both interfaces.

\paragraph{{Wavefunction matching.}}
Let $\psi^E_m(s)$ be the scattering wavefunction as given in previous
sections.
The current must be continuous at the interfaces. Employing the expression
\pref{eq:curu}, we arrive at the conditions
\begin{subequations}
\begin{align}
\psi^E_m(0+) &= \psi^E_m(0-)\\
-\iu \partial_s\psi^E_m(0+) &= -\iu\partial_s \psi^E_m(0-)\\
\psi^E_m(L-) &= \psi^E_m(L+)\\
-\iu \partial_s\psi^E_m(L-) &= -\iu\partial_s \psi^E_m(L+).
\end{align}
\label{eq:matching}
\end{subequations}
For a given direction of the incoming spin,
the wavefunction contains 8 unknown coefficients, which can be completely
determined from the \epref{eq:matching}.

In presence of a delta barrier at the left contact $(s=0)$,
the Hamiltonian gains the potential term
\[
v(s) = c\delta(s)\cdot \hat\sigma_0.
\]
The matching condition for the left contact becomes
\begin{align}
\psi^E_m(0+) &= \psi^E_m(0-)\\
-\iu \partial_s\psi^E_m(0+) &= -\iu\partial_s \psi^E_m(0-) - \iu 2c\,\psi^E_m(0).
\end{align}

\begin{figure}[p]
\begin{center}
\includegraphics[width=.7 \columnwidth]{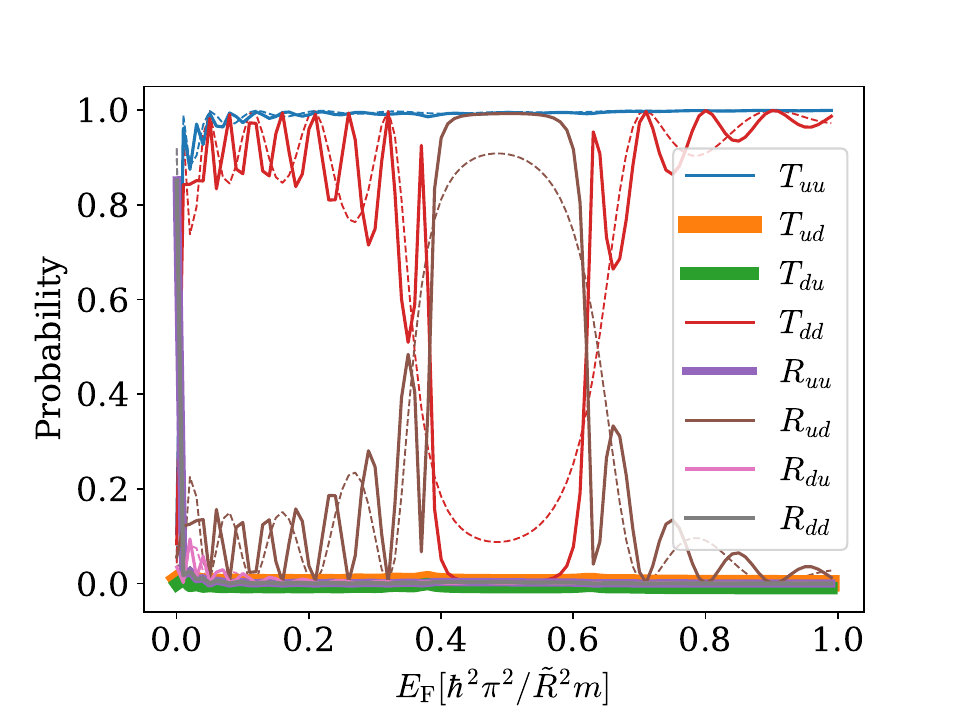}\\
\includegraphics[width=.7\columnwidth]{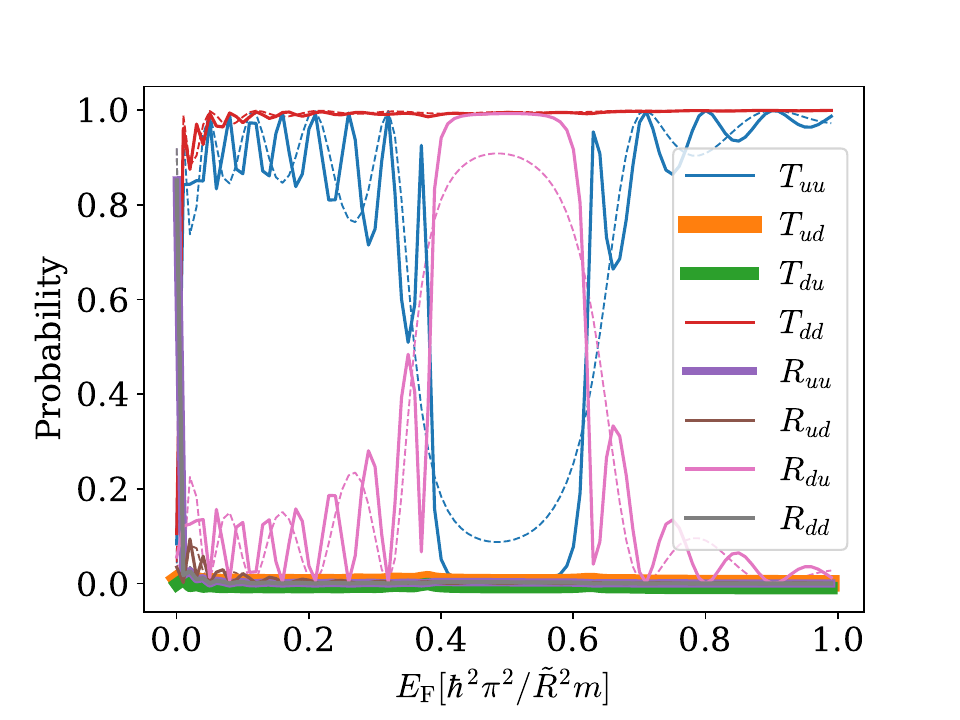}
\caption{\label{fig:sprob}Scattering probabilities for transmission and
reflection from the \textbf{left} (top panel) and \textbf{right} (bottom) leads,
 (u)p and (d)own spins. The probabilities are $m$-independent for $m=\pm 1$.
Parameters: $\tilde\kappa=0.1, \helc = - 1$,
same as in Fig.~3 of the main text, and the helix has 6 (solid lines) and 3 (dashed lines) turns. $E=0,0.4, 0.6$ mark the bottom
of the conduction band of the lead, and the limits of the gap of the helical
bandstructure.}
\end{center}
\end{figure}

\paragraph{{Scattering probabilities.}}
\figref{fig:sprob} shows the left and right transmission and reflection
probabilities
for a helix with 6 turns and the same parameters as in Fig.~3.

\section{\label{sec:symmetries}Symmetries of the scattering matrix}
\newcommand{\rrl}[1]{r^{(\Lld\Lld)}_{#1, \emm}}
\newcommand{\rrr}[1]{r^{(\Rld\Rld)}_{#1, \emm}}
\newcommand{\ttl}[1]{t^{(\Rld\Lld)}_{#1, \emm}}
\newcommand{\ttr}[1]{t^{(\Lld\Rld)}_{#1, \emm}}
\newcommand{\emm}{m}

We assume a scattering problem, defined by the scattering Hamiltonian
\begin{equation}
\label{eq:ham_ks}
\hat H_{m}(s) = 
 \half\left({ -\iu \dv{s}} -\helc\hat\sigma_z \right)^2 
- \helc\tilde\kappa(s) m\hat \sigma_y,
\end{equation}
where $\tilde\kappa(s) = \tilde\kappa$ for $-L/2 < s < L/2$ (central region) and zero in the left lead ($s<-L/2$) and the right lead $(s>L/2)$.
In what follows we shall make use of the basic invariance properties of $\hat H_{m}(s)$ in order to derive symmetry relations of the scattering matrix.
These invariance properties may hold true under more general conditions for a wider class of Hamiltonians. For example, the contacts may be modeled by a continuous decrease of $\tilde\kappa(s)$, or the
contacts could model vacuum tunneling.
We start our exposure with charge conservation (implying the well known unitarity), follow with time-reversal and end up with spatial inversions.

% We assume a scattering region, described by the Hamiltonian $\hat H$, that is connected
% to a pair of leads. The system allows for spin and orbital angular momenta, labeled
% by the projections on the $z$-axis, $\sigma = \sd,\su$ and $m$.

\subsection{Notation}
The leads ($\tilde\kappa(s)=0$) represent ``straight tubes'' and conserve both $\hat \sigma_z$ and the orbital angular momentum $m$. On the other hand, in the central region $-L/2 < s < L/2$ only $m$ is conserved.
An incoming wave in the left lead at energy $E$ with spin $\su$ in the
channel $m$
 is denoted by $\ket{\Phi_{E, \su, m}^{(\text{In},\Lld)}}$.
Similarly, an incoming wave in the right lead with the same quantum numbers
is denoted by $\ket{\Phi_{E, \su, m}^{(\text{In},\Rld)}}$.
Since the scattering processes conserve $E$ and $m$, for a fixed $m$ we can write
a general incoming wave as a 4-vector,
\begin{equation}
\ket{\Phi_{E,m}^{(\text{In})}} =
\left(
\ket{\Phi_{E, \su, m}^{(\text{In},\Lld)}},
\ket{\Phi_{E, \sd, m}^{(\text{In},\Lld)}},
\ket{\Phi_{E, \su, m}^{(\text{In},\Rld)}},
\ket{\Phi_{E, \sd, m}^{(\text{In},\Rld)}}
\right)^\mathrm T.
\end{equation}
The scattering operator $\hat S_m(E)$ maps $\ket{\Phi_{E,m}^{(\text{In})}}$ into
an outgoing wave $\ket{\Phi_{E,m}^{(\text{Out})}}$,
\begin{equation}
\ket{\Phi_{E,m}^{(\text{Out})}} = \hat S_m (E)\ket{\Phi_{E,m}^{(\text{In})}},
\end{equation}
where
\begin{equation}
\ket{\Phi_{E,m}^{(\text{Out})}} =
\left(
\ket{\Phi_{E, \su, m}^{(\text{Out},\Lld)}},
\ket{\Phi_{E, \sd, m}^{(\text{Out},\Lld)}},
\ket{\Phi_{E, \su, m}^{(\text{Out},\Rld)}},
\ket{\Phi_{E, \sd, m}^{(\text{Out},\Rld)}}
\right)^\mathrm T.
\end{equation}
From now on we suppress the energy dependence in the notation; we denote the matrix elements of $\hat S_m$ in this basis set by
\begin{equation}
\label{eq:selements}
\begingroup
\renewcommand*{\arraystretch}{2.0}
% your vertically stretched pmatrix expression
\hat S_m \equiv \begin{pmatrix}
\rrl{\su\su} & \rrl{\su\sd} & \ttr{\su\su} & \ttr{\su\sd} \\
\rrl{\sd\su} & \rrl{\sd\sd} & \ttr{\sd\su} & \ttr{\sd\sd} \\
\ttl{\su\su} & \ttl{\su\sd} & \rrr{\su\su} & \rrr{\su\sd} \\
\ttl{\sd\su} & \ttl{\sd\sd} & \rrr{\sd\su} & \rrr{\sd\sd}
\end{pmatrix} \equiv \begin{pmatrix}
\rrl{\sigma\sigma'} & \ttr{\sigma\sigma'} \\
\ttl{\sigma\sigma'} & \rrr{\sigma\sigma'}
\end{pmatrix}
\endgroup
\end{equation}

\subsection{Unitarity from the charge conservation}
Each matrix element of \pref{eq:selements}, when squared, gives a scattering
probability. We shall use capital letters for the probabilities, for instance,
$|\ttl{\su\su}|^2 = \Tl{\su}{\su}$. For reference, we reproduce here some
consequences of unitarity of the scattering matrix:
\begin{subequations}
\label{eq:unitarity}
\begin{align}
1 &= \sum_{\sigma} \Rl{\sigma}{\sigma'} + \Tl{\sigma}{\sigma'}\\
\label{eq:unit2}
1 &= \sum_{\sigma} \Rr{\sigma}{\sigma'} + \Tr{\sigma}{\sigma'}\\
\label{eq:unit3}
1 &= \sum_{\sigma'} \Rl{\sigma}{\sigma'} + \Tr{\sigma}{\sigma'}\\
1 &= \sum_{\sigma'} \Rr{\sigma}{\sigma'} + \Tl{\sigma}{\sigma'}
\label{eq:unit4}
\end{align}
\end{subequations}
The first two equations state that the incoming flux must be distributed
in a conserving way into the outgoing fluxes. The last two equations
require normalization of the outgoing waves.

We shall make frequent use of an auxiliary identity obtained
by multiplying the \espref{eq:unit3}{eq:unit4} by $\sigma$ and summing,
\begin{subequations}
\begin{align}
\sum_{\sigma\sigma'}\sigma\Trl{\sigma}{\sigma'} &=
-\sum_{\sigma\sigma'}\sigma\Rll{\sigma}{\sigma'},\\
\sum_{\sigma\sigma'}\sigma\Tl{\sigma}{\sigma'} &=
-\sum_{\sigma\sigma'}\sigma\Rr{\sigma}{\sigma'}
\end{align}
\label{eq:aux}
\end{subequations}

\subsection{Time-reversal invariance}
The condition of time reversal invariance imposes the following
restriction on the matrix elements:
\begin{equation}
\bra{\Phi_m^{(\text{Out})}} \hat S_m \ket{\Phi_m^{(\text{In})}}
=
\bra{T \Phi_m^{(\text{In})}} \hat S_m \ket{ T \Phi_m^{(\text{Out})}}
\end{equation}
$T=-\iu\hat\sigma_y K$ is the time reversal operator;
its effect on the In and Out waves amounts to
\begin{align}
\ket{T \Phi_m^{(\text{In})}} =
\left(
\ket{\Phi_{E, \sd,-m}^{(\text{Out},\Lld)}},
-\ket{\Phi_{E, \su,-m}^{(\text{Out},\Lld)}},
\ket{\Phi_{E, \sd,-m}^{(\text{Out},\Rld)}},
-\ket{\Phi_{E, \su,-m}^{(\text{Out},\Rld)}}
\right)^\mathrm T \\
\ket{T\Phi_m^{(\text{Out})}} =
\left(
\ket{\Phi_{E, \sd,-m}^{(\text{In},\Lld)}},
-\ket{\Phi_{E, \su,-m}^{(\text{In},\Lld)}},
\ket{\Phi_{E, \sd,-m}^{(\text{In},\Rld)}},
-\ket{\Phi_{E, \su,-m}^{(\text{In},\Rld)}}
\right)^\mathrm T.
\end{align}
Notice that the minus sign is specific to the transformation of spinors.
The following cross-relationships between matrix elements are implied,
\begin{subequations}
\label{eq:tri}
\begin{align}
r^{(\Lld\Lld)}_{\sigma,\sigma', m} &= (\sigma\sigma')\,r^{(\Lld\Lld)}_{-\sigma',-\sigma, -m},
&
R^{(\Lld\Lld)}_{\su\su, m} &= R^{(\Lld\Lld)}_{\sd\sd, -m}
\label{eq:tri_a}
\\
r^{(\Rld\Rld)}_{\sigma,\sigma', m} &= (\sigma\sigma')\,r^{(\Rld\Rld)}_{-\sigma',-\sigma, -m},
&
R^{(\Rld\Rld)}_{\su\su, m} &= R^{(\Rld\Rld)}_{\sd\sd, -m},
\\
t^{(\Lld\Rld)}_{\sigma,\sigma', m} &= (\sigma\sigma')\,t^{(\Rld\Lld)}_{-\sigma',-\sigma, -m},
&
T^{(\Lld\Rld)}_{\sigma,\sigma', m} &= T^{(\Rld\Lld)}_{-\sigma',-\sigma, -m}.
\label{eq:tri3}
\end{align}
\end{subequations}
The above formul\ae{} have important consequences for the occurrence of spin polarization. Consider an unpolarized incident flux in the left lead. The spin conserving reflection
processes are balanced due to \epref{eq:tri_a}. If spin polarization
results from the reflection, it must be due to spin flipping processes
\cite{Yang2019}.

If we set $m=0$ in the \epref{eq:tri} we restrict to the single channel.
Very strict conditions follow for the reflection amplitudes,
\begin{align}
r^{(\Lld\Lld)}_{\sigma\sigma} &= r^{(\Lld\Lld)}_{-\sigma,-\sigma}, & 
r^{(\Rld\Rld)}_{\sigma\sigma} &= r^{(\Rld\Rld)}_{-\sigma,-\sigma}\\
r^{(\Lld\Lld)}_{\sd\su} &= r^{(\Lld\Lld)}_{\su\sd} = 0, & 
r^{(\Rld\Rld)}_{\sd\su} &= r^{(\Rld\Rld)}_{\su\sd} = 0.
\end{align}
When combined with unitarity, these relations prohibit spin
filtering in single channel wires, as proven by Kiselev and Kim\cite{Kiselev2005}.

\subsection{Spatial symmetries}
Here we explore the consequences of various operations on the spatial
(spin and orbital) degrees of freedom.

\subsubsection{Spin inversion}
The basic algebraic identities $\qty[\hat\sigma_z]^3 = \hat\sigma_z$ and $\hat\sigma_z
\hat\sigma_y\hat\sigma_z= -\hat\sigma_y$ can be combined to yield the following
invariance property
\begin{equation}
\label{eq:parity0}
\hat\sigma_z \hat H_m(s) \hat\sigma_z = \hat H_{-m}(s)
\end{equation}
where $\hat H_m(s)$ is introduced in the \epref{eq:ham_ks}.\footnote{%
We remind that the operator $\hat\sigma_z = \iu\exp(-\iu\hat\sigma_z\pi/2)$
performs a rotation by $\pi$
around the $z$ axis of the spin degrees of freedom (up to a phase factor); it effectively inverts the sign
of $\hat\sigma_y$.
Such rotation can be compensated for by flipping the sign of $m$.
}.
The above relation holds for both the straight tube and the helix.
Therefore, the scattering matrix obeys
\begin{equation}
\hat\sigma_z \hat S_{-m} \hat\sigma_z = \hat S_{m}
\end{equation}
\ie
\begin{subequations}
\label{eq:parity}
\begin{align}
r^{(\Lld\Lld)}_{\sigma,\sigma', m} &= (\sigma\sigma')\,r^{(\Lld\Lld)}_{\sigma,\sigma', -m},
&
R^{(\Lld\Lld)}_{\sigma,\sigma', m} &= R^{(\Lld\Lld)}_{\sigma,\sigma', -m},
\\
r^{(\Rld\Rld)}_{\sigma,\sigma', m} &= (\sigma\sigma')\,r^{(\Rld\Rld)}_{\sigma,\sigma', -m},
&
R^{(\Rld\Rld)}_{\sigma,\sigma', m} &= R^{(\Rld\Rld)}_{\sigma,\sigma', -m}\\
t^{(\Lld\Rld)}_{\sigma,\sigma', m} &= (\sigma\sigma')\,t^{(\Lld\Rld)}_{\sigma,\sigma', -m},
&
T^{(\Lld\Rld)}_{\sigma,\sigma', m} &= T^{(\Lld\Rld)}_{\sigma,\sigma', -m}\\
t^{(\Rld\Lld)}_{\sigma,\sigma', m} &= (\sigma\sigma')\,t^{(\Rld\Lld)}_{\sigma,\sigma', -m},
&
T^{(\Rld\Lld)}_{\sigma,\sigma', m} &= T^{(\Rld\Lld)}_{\sigma,\sigma', -m}.
\end{align}
\end{subequations}
Clearly, all scattering probabilities are $m$-independent.

Combining the latter symmetry with time-reversal yields
identities equating elements of the scattering matrix from the same
$m$ sector, as can be derived using \espref{eq:parity}{eq:tri}:
\begin{subequations}
\label{eq:pt}
\begin{align}
r^{(\Lld\Lld)}_{\sigma,\sigma', m} &= r^{(\Lld\Lld)}_{-\sigma',-\sigma, m},
&
R^{(\Lld\Lld)}_{\su\su, m} &= R^{(\Lld\Lld)}_{\sd\sd, m}\\
r^{(\Rld\Rld)}_{\sigma,\sigma', m} &= r^{(\Rld\Rld)}_{-\sigma',-\sigma, m},
&
R^{(\Rld\Rld)}_{\su\su, m} &= R^{(\Rld\Rld)}_{\sd\sd, m}\\
t^{(\Lld\Rld)}_{\sigma,\sigma', m} &= t^{(\Rld\Lld)}_{-\sigma',-\sigma, m},
&
T^{(\Lld\Rld)}_{\sigma,\sigma', m} &= T^{(\Rld\Lld)}_{-\sigma',-\sigma, m}.
\label{eq:pt3}
\end{align}
\end{subequations}

\subsubsection{Parity}
The helicity of the helix is invariant under inversion of
all three spatial coordinates. Such an operation reverses the 
left and right leads, though. We exploit this property in what follows.

The Hamiltonian is not invariant under $s$-reversal $\hat P_s: s \mapsto -s$,
but under the joint operation of $\hat P_s$ and $\hat\sigma_y$,\cite{Sanchez2008}
\begin{equation}
\qty(\hat\sigma_y \hat P_s)\, \hat H_m(s)\, 
\qty(\hat P_s \hat\sigma_y) = \hat H_m(s),
\end{equation}
and so is $\hat S_m$. Consequently,
\begin{subequations}
\label{eq:parity2}
\begin{align}
r^{(\Lld\Lld)}_{\sigma,\sigma', m} &= (\sigma\sigma')\,r^{(\Rld\Rld)}_{-\sigma,-\sigma', m},
&
R^{(\Lld\Lld)}_{\sigma,\sigma', m} &= R^{(\Rld\Rld)}_{-\sigma,-\sigma', m},
\\
t^{(\Lld\Rld)}_{\sigma,\sigma', m} &= (\sigma\sigma')\,t^{(\Rld\Lld)}_{-\sigma,-\sigma', m},
&
T^{(\Lld\Rld)}_{\sigma,\sigma', m} &= T^{(\Rld\Lld)}_{-\sigma,-\sigma', m},
\label{eq:parity3}
\end{align}
\end{subequations}
where we used the fact that $\hat P_s$ interchanges the leads.
Combining parity and time reversal yields $T^{(\Lld\Rld)}_{\su\sd, m} = T^{(\Lld\Rld)}_{\sd\su, m}$
and $T^{(\Rld\Lld)}_{\su\sd, m} = T^{(\Rld\Lld)}_{\sd\su, m}$.

\subsubsection{Remark: Structure of the probability matrix}
The spatiotemporal invariance properties can be used to lay out the
structure of the
matrix of scattering probabilities, using 6 independent parameters only,
\begin{equation}
\begingroup
\renewcommand*{\arraystretch}{2.0}
% your vertically stretched pmatrix expression
\begin{pmatrix}
\Rl{\su}{\su} & \Rl{\su}{\sd} & \Tr{\su}{\su} & \Tr{\su}{\sd} \\
\Rl{\sd}{\su} & \Rl{\sd}{\sd} & \Tr{\sd}{\su} & \Tr{\sd}{\sd} \\
\Tl{\su}{\su} & \Tl{\su}{\sd} & \Rr{\su}{\su} & \Rr{\su}{\sd} \\
\Tl{\sd}{\su} & \Tl{\sd}{\sd} & \Rr{\sd}{\su} & \Rr{\sd}{\sd}
\end{pmatrix}
\endgroup
=
\begin{pmatrix}
a & b & c & d \\
g & a & d & f \\
f & d & a & g \\
d & c & b & a \\
\end{pmatrix}, \quad\text{independent of } m.
\end{equation}
We can recognize immediately that the symmetries of the scattering Hamiltonian
allow for spin polarization due to spin-flipping reflections 
and non-flipping transmissions. Specifically, spin polarization
is generated if the expressions
\[
\Rr{\sd}{\su} - \Rr{\su}{\sd},\ \Tl{\su}{\su}-\Tl{\sd}{\sd},\quad
\text{and with } \Rld \longleftrightarrow \Lld
\]
are non-zero. Other spin processes are balanced
and do not contribute to the polarization. 

%%%%%%%%%%%%%%%%%%%%%%%%%%%%%%%%%%%%%%%%%%%%%%%%%%%%%%%%%%%%%%%%
\section{\label{sec:landauer}Transport coefficients in the Landauer formalism and 
their symmetry relations}
We derive explicit expressions for the transport coefficients, \epref{eq:linearresponse},
in Landauer formalism, employing the properties of the scattering
probability matrix that we have given in the previous section.

\subsection{Expressions for the currents}
Charge current in the left lead is simply given by summing
(a) an incident flux in the left lead, (b) a back-reflected flux
in the left lead and (c) the fluxes transmitted from the right lead,
namely,
\begin{equation}
I^{(\Lld)} = \frac eh \sum_m\sum_{\sigma'} \int \mathrm dE
\left\{ \nflsp \left[1 - \sum_{\sigma}\Rl{\sigma}{\sigma'}\right] - \nfrsp \sum_{\sigma}\Tr{\sigma}{\sigma'}
\right\}.
\end{equation}
With
\begin{align*}
\mu^{\sigma'}_{\Lld}\ &=\ \EF + \half eV + \frac \hbar 2 W_z^{(\Lld)}\sigma'\\
\mu^{\sigma'}_{\Rld}\ &=\ \EF - \half eV + \frac \hbar 2 W_z^{(\Rld)}\sigma'
\end{align*}
[see also \epref{eq:imbalances}]
we Taylor-expand
the Fermi functions around equilibrium (at infinitesimal temperature) in order to
obtain the current in linear
response,
\begin{align}
I &= \frac eh \sum_m\sum_{\sigma'} \int \mathrm dE
\Biggl\{ n_\mathrm F(E-\EF) \qty[1 - \sum_\sigma \Rl{\sigma}{\sigma'} - \sum_\sigma \Tr{\sigma}{\sigma'}]\ +\\
&+\ \phantom{\frac\hbar2 \sigma'}\qty[\pdv{n_\mathrm F(E-\EF)}{\EF}]_{E,T}\cdot\qty[1 - \sum_\sigma \Rl{\sigma}{\sigma'} + \sum_\sigma \Tr{\sigma}{\sigma'}]\
\frac{eV}{2}\ +\\
&+\ \frac\hbar2 \sigma'\qty[\pdv{n_\mathrm F(E-\EF)}{\EF}]_{E,T}\cdot\qty[\qty(1 - \sum_\sigma \Rl{\sigma}{\sigma'})
 W_z^{(\Lld)} - \sum_\sigma \Tr{\sigma}{\sigma'} W_z^{(\Rld)}]
\Biggr\}.
\end{align}
Expressing the reflection coefficients through transmissions by using unitarity, \epref{eq:unitarity}, results in
\begin{itemize}
\item vanishing of the equilibrium term;
\item  furthermore, at zero temperature we get Landauer-B\"uttiker formula for the charge conductance, as given in the main text
\item coefficients of the charge-current response to spin accumulations,
\begin{subequations}
\begin{align}
\tilde G_z^{(\Lld)} &= +\frac e{4\pi} \sum_m\sum_{\sigma\sigma'} \sigma' \Tl{\sigma}{\sigma'}\\
\tilde G_z^{(\Rld)} &= -\frac e{4\pi} \sum_m\sum_{\sigma\sigma'} \sigma' \Tr{\sigma}{\sigma'}.
\end{align}
\label{eq:gtildes}
\end{subequations}
\end{itemize}

In a similar way it is possible to derive expressions for the spin currents,
\begin{align}
I_z^{(\Lld)} &= \frac 1{4\pi} \sum_m\sum_{\sigma'} \int \mathrm dE
(-1)\left\{ \nflsp \sum_{\sigma}\sigma\Rl{\sigma}{\sigma'} +  \nfrsp \sum_{\sigma}\sigma\Tr{\sigma}{\sigma'}
\right\}\\
I_z^{(\Rld)} &= \frac 1{4\pi} \sum_m\sum_{\sigma'} \int \mathrm dE
(+1)\left\{ \nfrsp \sum_{\sigma}\sigma\Rr{\sigma}{\sigma'} +  \nflsp \sum_{\sigma}\sigma\Tl{\sigma}{\sigma'}
\right\}.
\end{align}
Notice that the sign change from - to + is due to our definition of the positive direction
of the current, see the definitions in the main text. The linearized expressions
\begin{align*}
I_z^{(\Lld)} &= -\frac 1{4\pi} \sum_m\sum_{\sigma'} \int \mathrm dE
\Biggl\{ n_\mathrm F(E-\EF) \sum_\sigma\sigma  \qty[\Rl{\sigma}{\sigma'} + \Tr{\sigma}{\sigma'}]\\
&+ \qty(\pdv{n_\mathrm F}{\EF})\frac{eV}2\sum_\sigma\sigma  \qty[\Rl{\sigma}{\sigma'} - \Tr{\sigma}{\sigma'}]
 +\\
&+ \qty(\pdv{n_\mathrm F}{\EF}) \frac\hbar2 \sum_\sigma\sigma\sigma' \qty[\Rl{\sigma}{\sigma'}
 W_z^{(\Lld)} + \Tr{\sigma}{\sigma'} W_z^{(\Rld)}]
\Biggr\},
\end{align*}
\begin{align*}
I_z^{(\Rld)} &= +\frac 1{4\pi} \sum_m\sum_{\sigma'} \int \mathrm dE
\Biggl\{ n_\mathrm F(E-\EF)\sum_\sigma\sigma \qty[ \Rr{\sigma}{\sigma'} + \Tl{\sigma}{\sigma'}]\\
&+ \qty(\pdv{n_\mathrm F}{\EF})\frac{eV}{2}\sum_\sigma\sigma \qty[ - \Rr{\sigma}{\sigma'} + \Tl{\sigma}{\sigma'}]
 +\\
&+ \qty(\pdv{n_\mathrm F}{\EF})\frac\hbar2 \sum_\sigma\sigma'\sigma \qty[ \Rr{\sigma}{\sigma'})
 W_z^{(\Rld)} + \Tl{\sigma}{\sigma'} W_z^{(\Lld)}]
\Biggr\}
\end{align*}
deliver using \epref{eq:aux}
\begin{itemize}
\item vanishing equilibrium spin currents on accounts of unitarity,
\item spin conductances as given in the main text, Eqs.~(\eqspins),
\item response of the spin currents to spin accumulations characterised by coefficients:
\begin{align*}
\coefzz^{(\Lld\Lld)} &= -\frac\hbar{8\pi} \sum_{\sigma\sigma'm} \sigma\sigma'\, \Rl{\sigma}{\sigma'}\\
\coefzz^{(\Lld\Rld)} &= -\frac\hbar{8\pi} \sum_{\sigma\sigma'm} \sigma\sigma'\, \Tr{\sigma}{\sigma'}\\
\coefzz^{(\Rld\Lld)} &= +\frac\hbar{8\pi} \sum_{\sigma\sigma'm} \sigma\sigma'\, \Tl{\sigma}{\sigma'}\\
\coefzz^{(\Rld\Rld)} &= +\frac\hbar{8\pi} \sum_{\sigma\sigma'm} \sigma\sigma'\, \Rr{\sigma}{\sigma'}.
\end{align*}
\end{itemize}

\subsection{Onsager's reciprocity relations}
The reciprocity of spin conductances, $G_z^{(\Lld,\Rld)}$, and the
coefficients, $\tilde G_z^{(\Lld,\Rld)}$, can be demonstrated readily.
Upon employing the \epref{eq:tri3} in the formul\ae{} \epref{eq:gtildes}
and comparing with the Eqs.~(\eqspins) of the main text we
readily confirm the reciprocal relations, \espref{eq:reciprocity1}{eq:reciprocity2}.
In the same way the reciprocity \pref{eq:reciprocity3} follows by employing
time reversal symmetry of transmission probabilities.

\subsection{Nonzero spin conductance is due to spin-flip reflections}
The spin conductances were defined in the main text.
Using \epref{eq:aux} we can express them using reflection coefficients only.
The spin-diagonal elements can be eliminated with the \epref{eq:tri_a} and
in time-reversal invariant situations only the spin-flipping processes
contribute,
\begin{subequations}
\begin{align}
G_z^{(\Rld)} &= \spinq\sum_m 
\qty(\Rr{\sd}{\su} - \Rr{\su}{\sd})\\
G_z^{(\Lld)} &= \spinq\sum_m 
\qty(\Rl{\sd}{\su} - \Rl{\su}{\sd}).
\end{align}
\label{eq:gzr}
\end{subequations}

\subsection{A proof of $G_z^{(\Rld)} = - G_z^{(\Lld)}$}
By adding the \epref{eq:gzr}
we obtain zero because of parity, \epref{eq:parity2}. It follows that
\begin{equation}
 G_z^{(\Rld)} = -G_z^{(\Lld)}.
\label{eq:main_result}
\end{equation}

Since the proof requires only basic symmetries, it is conceivable  
that the relation \pref{eq:main_result} holds true for a wider
class of Hamiltonians, the \epref{eq:ham_ks} being an example.
For instance, the contacts might be modeled by a smooth
behavior of $\tilde\kappa(s)$ or the contacts might involve barrier tunneling.
However, once
the inversion symmetry is lost, \ie{} when the coupling to the left lead
is not the same as to the right, the identity \epref{eq:main_result} ceases to hold.

It is also possible to derive the identity \epref{eq:main_result}
by using transmission probabilities only. To this end, we add
 the expressions
\begin{subequations}
\begin{align}
G_z^{(\Rld)} &= \spinq\sum_m 
\qty(
\Tll{\su}{\su} + \Tll{\su}{\sd} - \Tll{\sd}{\su} -  \Tll{\sd}{\sd})\\
G_z^{(\Lld)} &= \spinq\sum_m 
\qty(
\Trl{\su}{\su} + \Trl{\su}{\sd} - \Trl{\sd}{\su} -  \Trl{\sd}{\sd}).
\end{align}
\end{subequations}
Discarding the constant prefactor we
get an expression that can be regrouped into four rounded brackets,
\[
\sum_m\qty[
\qty( \Tll{\su}{\su} - \Trl{\sd}{\sd} ) +
\qty( \Trl{\su}{\su} - \Tll{\sd}{\sd} ) +
\qty( \Trl{\su}{\sd} - \Tll{\sd}{\su} ) +
\qty( \Tll{\su}{\sd} - \Trl{\sd}{\su} ) ].
\]
The first two brackets vanish on account of \epref{eq:tri3} and the remaining two
because of \epref{eq:parity3}.

\section{Spin and charge densities in the leads}
Equilibrium spin and charge densities and their bias-induced
counterparts are elaborated in this section from scattering
wavefunctions. For the sake of specificity $m=\pm 1$.

\subsection{Local densities of scattering states}
\paragraph{Wavefunctions in the left lead.}
At a fixed energy $E = \half q^2$ and $m$ there are 4 scattering states,
mapping asymptotically to spin up/down particles incoming from the left
and right lead. In this section we label them L1, L2, R1 and R2.
Their spinor wavefunctions were given in Sec.~\ref{sec:eval} and
here we reproduce their expressions in the left lead ($s<0$)
\begin{subequations}
\label{eq:leftwfn}
\begin{align}
\psi^E_{L1,m}(s) &= \fsp\qty[ e^{\iu qs} \ket{\su} +
    e^{-\iu qs} \sum_\sigma r^{(\Lld\Lld)}_{\sigma\su,m}\ket{\sigma}] \\
\psi^E_{L2,m}(s) &= \fsp\qty[ e^{\iu qs} \ket{\sd} +
    e^{-\iu qs} \sum_\sigma r^{(\Lld\Lld)}_{\sigma\sd,m}\ket{\sigma}] \\
\psi^E_{R1,m}(s) &= \fsp e^{-\iu qs}\sum_\sigma t^{(\Lld\Rld)}_{\sigma\su,m}\ket{\sigma} \\
\psi^E_{R2,m}(s) &= \fsp e^{-\iu qs}\sum_\sigma t^{(\Lld\Rld)}_{\sigma\sd,m}\ket{\sigma}.
\end{align}
\end{subequations}
Notice that we employ a plane-wave normalization factor $1/\sqrt{2\pi}$, unlike in Sec.~\ref{sec:eval}.

\begin{comment}
\paragraph{Local densities.}
For a given $q>0$, the spin-resolved densities from an incoherent contribution
of the four states, \epref{eq:leftwfn}, read
\begin{align*}
n_\su(q, s) &=
\frac 1{2\pi}\sum_m \qty{ \left| e^{\iu qs} + e^{-\iu qs} r^{(\Lld\Lld)}_{\su\su,m}\right|^2
+ \Rl{\su}{\sd} + \Tr{\su}{\sd} + \Tr{\su}{\su} } \\
&= \frac 1{\pi} \sum_m \Re \qty{e^{-\iu2qs}\, r^{(\Lld\Lld)}_{\su\su,m}+
1},\\
n_\sd(q, s)  &=
\frac 1{2\pi}\sum_m \qty{ \left| e^{\iu qs} + e^{-\iu qs} r^{(\Lld\Lld)}_{\sd\sd,m}\right|^2
+ \Rl{\sd}{\su} + \Tr{\sd}{\su} + \Tr{\sd}{\sd} } \\
&= \frac 1{\pi} \sum_m \Re\qty{ e^{-\iu2qs}\, r^{(\Lld\Lld)}_{\sd\sd,m} +
1},
\end{align*}
where we have taken advantage of unitarity, \epref{eq:unit3}, to simplify the expressions.

\end{comment}

\paragraph{Direction-resolved densities.}
In equilibrium, spin and charge densities can be obtained straightforwardly
by calculating the local density of states from the above wavefunctions.
In bias-induced stationary nonequilibrium we need to discriminate 
between states incoming from left and right.

For given $q>0$ and $s<0$,
the spin- and direction-resolved densities from 
the states \pref{eq:leftwfn} read
\begin{align*}
n_{\su,L}(q, s) &= \frac 1{2\pi}\sum_m\qty{ 1 + 2\Re\qty[ e^{-\iu 2 qs} r^{(\Lld\Lld)}_{\su\su,m} ]
+ \Rl{\su}{\sd} + \Rl{\su}{\su} },\\
n_{\sd,L}(q, s) &= \frac 1{2\pi}\sum_m\qty{ 1 + 2\Re\qty[ e^{-\iu 2 qs} r^{(\Lld\Lld)}_{\sd\sd,m} ]
+ \Rl{\sd}{\sd} + \Rl{\sd}{\su} },\\
n_{\su,R}(q, s) &= \frac 1{2\pi}\sum_m\qty{ \Tr{\su}{\sd} + \Tr{\su}{\su} },\\
n_{\sd,R}(q, s) &= \frac 1{2\pi}\sum_m\qty{ \Tr{\sd}{\sd} + \Tr{\sd}{\su} }.
\end{align*}

\paragraph{Spin and charge densities at energy
$E=\half q^2$.}
The (dimensionless) charge and spin densities follow directly from the above relations by summing the L and R contributions equally. First,
the charge density at $s<0$ reads
\begin{align}
n_0 (q, s) &=\sum_\sigma\phantom\sigma
\qty[ n_{\sigma,L}(q, s) + n_{\sigma,R}(q, s)] = 
\frac 1{\pi} \sum_m \Re\qty{2 + e^{-\iu2qs}\,\qty( r^{(\Lld\Lld)}_{\su\su,m} + r^{(\Lld\Lld)}_{\sd\sd,m})},
\label{eq:cseq}
\end{align}
where \epref{eq:unitarity} was applied.
Note that the charge density from eigenstates at energy $E$
in a clean infinite tube is $8/{2\pi}$ (due to twofold spin, orbital
and direction degeneracies), which is just the first term
in the above expression. The oscillating term occurs because (only) the spin-diagonal
reflection amplitudes lead to a \textit{coherent} sum
of two planewaves with opposite
momenta. Spin-flipping processes contribute incoherently, because
the incident and reflected waves pertain to disjoint
Hilbert subspaces
due to the opposite spins of the waves.

Second, the spin density vanishes on accounts
of time-reversal invariance (TRI),
\epref{eq:tri_a}.
\begin{align}
n_z(q, s) &=\sum_\sigma\sigma\qty[ n_{\sigma,L}(q, s) + n_{\sigma,R}(q, s)] =
\frac 1{\pi} \sum_m \Re \qty{e^{-\iu2qs}\,\qty( r^{(\Lld\Lld)}_{\su\su,m} - r^{(\Lld\Lld)}_{\sd\sd,m})}
\ \overset{\text{\tiny(TRI)}}{=} \ 0.
\end{align}

\paragraph{Spin and charge densities at energy $E$ for nonequilibrium.}
At a finite bias voltage, the symmetry between L and R states is broken. The charge
density is determined by the difference between L and R densities provided that
the bias difference $V$ applies symmetrically on both leads.
In such cases, the difference reads
\begin{multline}
\sum_\sigma \qty[n_{\sigma,L} (q, s) - n_{\sigma,R} (q, s)] =\\
= \frac 1{2\pi}\sum_m\qty{2 + 2\Re\qty[ e^{-\iu 2 qs}
\qty( r^{(\Lld\Lld)}_{\su\su,m} + r^{(\Lld\Lld)}_{\sd\sd,m}) ]
+\sum_{\sigma\sigma'}\qty( \Rl\sigma{\sigma'} - \Tr\sigma{\sigma'})}.
\end{multline}
The terms above have a rather transparent meaning. 
Employing the \epref{eq:unitarity} to convert the reflection probabilities into transmission
and recognizing the linear conductance formula, the previous expression
simplifies to
\begin{align}
\nonumber  &\phantom{=} \frac 1{\pi}\sum_m\qty{2 + \Re\qty[ e^{-\iu 2 qs}
\qty( r^{(\Lld\Lld)}_{\su\su,m} + r^{(\Lld\Lld)}_{\sd\sd,m}) ]
-\sum_{\sigma\sigma'} \Tr\sigma{\sigma'}}\\
\text{\small(TRI)}\quad
      &= \frac 1{\pi}\qty(4 - \frac{h}{e^2}G) + \frac 2{\pi}\sum_m \Re\qty[ e^{-\iu 2 qs}
 r^{(\Lld\Lld)}_{\su\su,m}].
\label{eq:cnoneq}
\end{align}
The application of TRI in the last line is indicated.
The expression \pref{eq:cnoneq} offers an interpretation:
The total number of L fluxes is 4. To this number, an
effective count of backreflected states, $-hG/e^2$, and
an interference term are to be added.

Analogously for the spin density in the left lead,
\begin{multline*}
\sum_\sigma \sigma\qty[n_{\sigma,L} (q, s) - n_{\sigma,R} (q, s)] =\\
= \frac 1{\pi}\sum_m\qty{\Re\qty[ e^{-\iu 2 qs}
\qty( r^{(\Lld\Lld)}_{\su\su,m} - r^{(\Lld\Lld)}_{\sd\sd,m}) ]
+\sum_{\sigma\sigma'}\sigma \qty(\Rl\sigma{\sigma'} - \Tr\sigma{\sigma'})}.
\end{multline*}
Unitarity, TRI and the definition of spin conductance simplify the expression,
\begin{align}\nonumber
\sum_\sigma \qty[n_{\sigma,L} (q, s) - n_{\sigma,R} (q, s)]
               &= -\frac 1{\pi}\sum_m \sum_{\sigma\sigma'}\sigma \Tr\sigma{\sigma'}\\
               &= -\frac 1{\pi}\frac{4\pi}e\, G_z^{(\Lld)}\Biggr|_{E={q^2}/2}.
\label{eq:snoneq}
\end{align}

\subsection{Surface- and bulk-accumulated densities} 
In order to get energy-unresolved
equilibrium and non-equilibrium charge and spin
densities (accumulations), the formul\ae{} (\ref{eq:snoneq},\ref{eq:cnoneq},\ref{eq:cseq})
have to be integrated over an energy window.

\paragraph{Equilibrium densities.}
The integration of the \epref{eq:cseq} delivers a homogeneous
term and a term that oscillates,
\[
n_0(s)\Bigr|_{V=0}
= n_\infty\,  +\, \frac 2\pi\sum_m\int_0^{q_\mathrm F}\frac{\mathrm dq}{2\pi}\,
\Re\qty[ e^{-\iu 2qs}\,r_{\su\su,m}^{(\Lld\Lld)}],
\quad \qF = \sqrt{2\EF}, \ s<0.
\]
with $n_\infty = 2\qF/\pi^2$ being the Fermi ground-state density
of a transparent infinite four-channel conductor.
The oscillating term describes familiar Friedel oscillations that decay asymptotically
as $1/\qF s$ from the contact into the bulk. The equilibrium charge density
therefore has a bulk term and a \textit{surface
accumulation}.

The spin density in equilibrium is zero because of TRI.

\paragraph{Nonequilibrium densities induced by a voltage bias.}
These observables are given by summing the density of L states
with energies $E\in\langle \EF, \EF+V/2\rangle$
and subtracting the contribution of R states in $\langle \EF -V/2, \EF\rangle$.
We achieve this readily by integrating the \epref{eq:cnoneq}
in
$q\in\langle \qF, \sqrt{2(\EF + V/2)}\rangle$. In first order in $V$
the charge density for $s<0$ reads
\begin{align*}
\delta n_0(s) &= \int_{\qF}^{\qF+ V/2\qF}\frac{\mathrm dq}{2\pi}
\qty{ \frac{1}{\pi} \qty(4 - \frac{h}{e^2}G)
+
 \frac 2\pi\sum_m\, 
\Re\qty[e^{-\iu 2qs}\,r_{\su\su,m}^{(\Lld\Lld)}]}
\ &{+}\,  \mathcal O\qty( V^2 )\\
&= V\, \varrho_\mathrm L(\EF) \qty[\qty(4 - \frac{h}{e^2}G) + 
2\sum_m \Re{e^{-\iu 2qs} r_{\su\su,m}^{(\Lld\Lld)}}] 
\ &{+}\,  \mathcal O\qty( V^2 ),
\end{align*} 
where we introduced the local density of left-moving states in
 a single channel (spinless) transparent 
wire 
\begin{equation}
\label{eq:ldosl}
\varrho_\mathrm L(\EF) = \frac 1     {4\pi^2\qF} =
 \frac 18 \left.\pdv{n_\infty}{E}\right|_{E=\EF}.
\end{equation}

The bulk term $V \varrho_\mathrm L(\EF)\qty(4 - \frac{h}{e^2}G)$ vanishes in the transparency limit. The oscillating term
is straightforwardly connected to the equilibrium Friedel oscillations.
\begin{comment}
Beyond linear response, the oscillating term
\[
\sum_m \int_{\qF}^{\sqrt{2(\EF + V/2)}}
\Re{e^{-\iu 2qs} r_{\su\su,m}^{(\Lld\Lld)}}
\,\frac{\mathrm dq}{\pi^2}
\]
yields a decaying Friedel oscillation,
\ie{} a voltage-induced surface charge.
\end{comment}

The spin-density is nonzero because the voltage breaks symmetry
between the L and R states; from the \epref{eq:snoneq} we get
\begin{equation}
\label{eq:dnz}
\delta n_z(s) = -V\, \varrho_\mathrm L(\EF) \frac{4\pi}e G_z^{(\Lld)}
\, +\, \mathcal O\qty(V^2),\quad s<0.
\end{equation}
The voltage-induced spin density does not have a surface component, only a bulk one.

\paragraph{Spin accumulation.}
The homogeneous voltage-induced spin density offers a simple estimate for
the spin accumulation, \ie{} the spin-polarized chemical potential
(see \epref{eq:imbalances}), in linear response to $V$.
Namely,
\begin{align}
\hbar \,\saW^{(\Lld)}(V) = \mu_\su^{(\Lld)}(V) - \mu_\sd^{(\Lld)}(V)
\ & \approx\
\left.\pdv{ \EF }{ n } \right|_{n\,=\,n_\infty} \delta n_z(s) \\
&= -V\, \frac{\pi}{2e} G_z^{(\Lld)},
\end{align}
where we used (\ref{eq:dnz},\ref{eq:ldosl}).

The expression holds true only if one can neglect spin scattering at the
 interface between the lead (tube) and 
the contacts of the battery.

\bibliography{si-references}